\documentclass[
reprint,
groupedaddress,
amsmath,
amssymb,
aps,
floatfix,
]{revtex4-2}

\usepackage{graphicx}% Include figure files
\usepackage{dcolumn}% Align table columns on decimal point
\usepackage{bm}% bold math
\usepackage{float}

\usepackage{natbib}[sort&compress]
\usepackage{xcolor}

\usepackage{color}
\usepackage{soul}
\usepackage{amssymb, nicefrac}
\usepackage[most]{tcolorbox}

\usepackage{hyperref}
\hypersetup{
    colorlinks=true,
    linkcolor=blue,
    filecolor=magenta,      
    urlcolor=blue,
    citecolor=blue,
    pdfpagemode=FullScreen,
    }

\usepackage{cleveref}

\begin{document}

%\preprint{}

\title{The influence of the boundary conditions on characteristics of nuclear fission}

\author{Pavel V. Kostryukov}
\email[]{E-mail: kostr@kft.umcs.lublin.pl}
%\thanks{}
\author{Artur Dobrowolski}
\affiliation{Theoretical Physics Department, Maria Curie-Sklodowska University, 20-031 Lublin,
 Poland}

\date{\today}

\begin{abstract}

In this paper, using a quasi-classical statistical approach based on the Langevin equation, we simulate the fission dynamics of selected even-even $\rm U$, $\rm Pu$, $\rm Cm$, $\rm Cf$ and $\rm Fm$
actinide nuclei. As a preparatory part of the work, before solving the Langevin equations, the determination of transport parameters such as inertia and friction tensors within the hydrodynamic approach is performed. Potential energy surfaces are calculated within a macroscopic-microscopic approach in a three-dimensional space of deformation parameters defined within the Fourier decomposition of the surface radius function in cylindrical coordinates. Using the Lublin-Strasbourg drop model, Strutinsky shell correction and BCS-like pairing energy model with the projection onto good particle number, we calculate the nuclear total potential energy surfaces (PES). The restoration of the particle number in the superfluid approach is realized within the Generator Coordinate Method (GCM) with the so called Gaussian Overlap Approximation (GOA). The final study is concerned with the effect of the starting point of the stochastic Langevin trajectory on its time evolution and, more importantly, the conditions for judging whether such a trajectory for a given time moment describes an already passed fission nucleus or not. Collecting a large number of such stochastic trajectories allows us to assess the resulting fragment mass distributions, which appear to be in good agreement with their experimental counterparts for light and intermediate actinides. More serious discrepancies are observed for single isotopes of californium and fermium.

\end{abstract}

%\keywords{}

\maketitle

\vspace{-12pt}
\section{Introduction}
\vspace{-6pt}

This year marks the $85^{th}$ anniversary of the discovery that heavy atomic nuclei are not only radioactive but also can decay into fragments of variable mass numbers called later as fission process. Although fission has been extensively investigated over this long period, we still need to gain complete knowledge about this process. Of course, there have been several successful attempts at its theoretical description, leading to some combinations of various well-known macroscopic liquid drop-like models and microscopic shell and pairing corrections, realized usually by the Strutinsky and the BCS-like models, respectively (see, e.g., Refs~\cite{myers1966,strutinsky1967,strutinsky1968,brack1972,moller1988,moller2016,jachimowicz2017}) providing a correct description of fission characteristics, such as distributions of masses, charges, kinetic energies, a multiplicity of emitted particles, etc. Nevertheless, "white spots" still exist in the description of the dynamics of the studied phenomenon, especially at its last stage, when the fissile system is close to splitting into fragments.

This study aims to shed some light on the still persistent problems of the dynamical description of low-energy fission of atomic nuclei, knowing that the nature of the fission phenomenon is, to some extent, stochastic. The starting point of the discussion is constructing a model based on the well-known macroscopic-microscopic approach~\cite{strutinsky1967, strutinsky1968}, where the potential energy function is expressed via the collective degrees of freedom, known as deformation parameters of the nuclear surface. 

The nuclear surface geometry is defined by the so-called shape parametrization, which is given here as a Fourier expansion of the square of the distance of a given point on the surface to the symmetry axis, $\rho^2(z,\varphi)$. The amplitudes of such a linear combination standing in front of the sine and cosine functions are related to the deformation parameters of the PES ~\cite{schmitt2017}. 
The fission dynamics, where the temporal evolution of the surface shape is governed by the system of Langevin equations~\cite{abe1986}, is described by a set of classical Hamilton-like trajectories, taking into account the excitation energy, friction between moving nucleons, and diffusion effects. Particular attention is paid to investigating the initial and the trajectory-termination conditions, which are crucial in obtaining a reasonable agreement of the generated fragment mass distributions (FMD) of primary fission fragments with the empirical data. The model has been ''calibrated'' in order to characterize in the best possible way the induced by thermal and 15 MeV neutrons fission of $\rm ^{235}U$ nucleus. Afterward, with further minor generalizations, it has been applied to simulate the spontaneous and induced fission of composite even-even actinides with proton number $Z$ in the region of 92-100.

The work has the following structure: after the introduction, the second chapter is devoted to the main points of the model. In the third chapter, we investigate the dependence of evaluated distributions of the primary fission fragments on the initial and termination conditions of the Langevin trajectories. In the fourth chapter, we apply the here fixed model to other than $\rm ^{236}U$ even-even nuclei and discuss the quality of our results by comparing them to the existing empirical data. We conclude our results in the last chapter.

\vspace{-6pt}
\section{Quasi-classical stochastic Langevin approach}
\vspace{-6pt}

The exact determination of the relevant fission process deformation parameters and the collective inertia and friction tensors are essential steps to successfully apply the Langevin approach to the evolution of a nucleus towards fission. Therefore, the critical issue of this kind of quasi-stochastic model is to obtain the change of nuclear surface shape with time, thus determining the set of a large number of trajectories $\mathbf{q(t)} = \{q_1(t),..., q_n(t)\}$ in the admitted $n$-dimensional deformation space.
 
\begin{figure}
\centering
    \includegraphics[width=\linewidth]{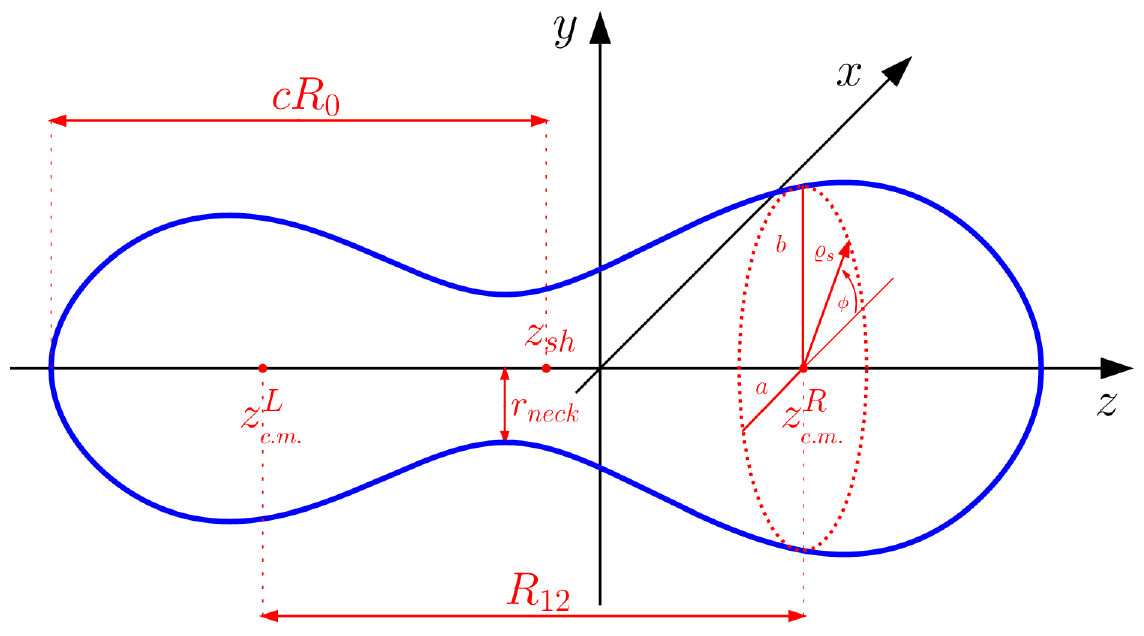}
    \vspace{-10pt}
    \caption[]{An example of the elongated nuclear surface obtained in the Fourier parameterization \eqref{eq:rho2}}\label{f01}
\end{figure}

\begin{figure*}
    \centering
    \includegraphics[width=\linewidth]{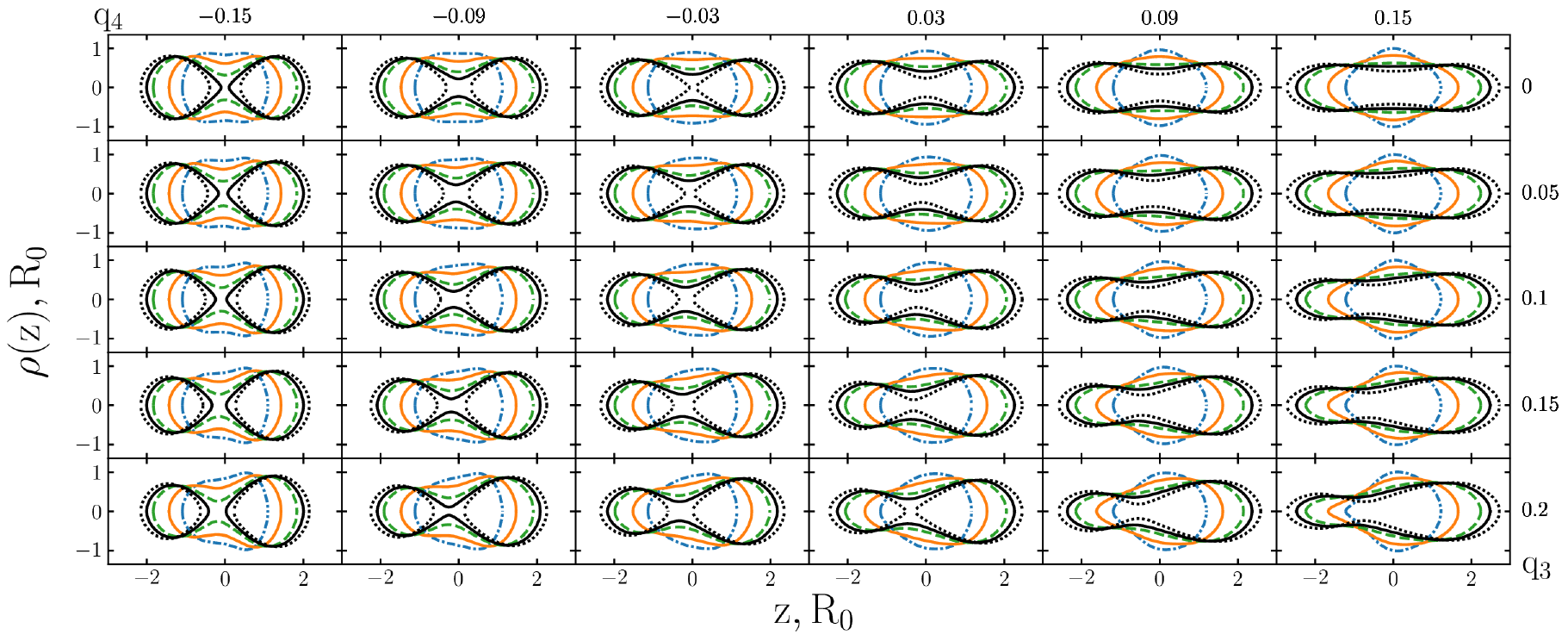}
    \vspace{-10pt}
    \caption[]{Fourier-like nuclear shapes for  $q_2 = 0.3$ (blue dot-dashed line), $q_2 = 1.0$ (green dashed line), $q_2 = 1.9$ (orange solid line), $q_2 = 2.35$ (black solid line), $q_2 = 2.9$ (black dotted line).}\label{f02}
\end{figure*}

At present, there exist various nuclear shape parametrizations, among which the most popular are spherical-harmonic decomposition \cite{trentalange1980}, Cassini ovaloids \cite{pashkevich_asymmetric_1971, adeev1971}, Funny-Hills and its later variations \cite{brack1972, pomorski_bartel2006} or two-center parameterization \cite{maruhn1972}. Nevertheless, in this paper one uses a relatively new, efficient, and fast convergent parametrization \cite{schmitt2017}, which represents the axially symmetric nuclear surface in cylindrical coordinates, $\rho_s^2(z, \mathbf{q})$ as a Fourier expansion of the form:
\begin{equation} \label{eq:rho2}
    \begin{split}
        \rho_s^2(z, \mathbf{q}) = R_0^2 \sum_{n=1} & \bigg[a_{2n}(\mathbf{q}) \cos{\left( \frac{2n-1}{2} \pi \frac{z - z_{sh}}{z_0} \right)} \\
        &+ \ a_{2n+1}(\mathbf{q}) \sin{\left(\frac{2n}{2} \pi \frac{z - z_{sh}}{z_0} \right)} \bigg],
    \end{split}
\end{equation}
where $R_0 = 1.2 \cdot A^{1/3}$ is the radius of the corresponding spherical nucleus, $z_{sh}$ is the displacement of the center of mass of the nucleus when $q_{2n+1}\neq 0 $ are considered. The dimensionless parameter $c$ is responsible for elongating the nuclear body along the $z$-axis. If $c>1$, nuclear shapes are {\it prolate} whereas $c<1$ produces oblate shapes. Therefore, the length of the nucleus measured along the $z$-axis is $2z_0 = 2cR_0$ where $\pm z_0$ determines the position of the right and the left end of the nucleus case $z_{sh}=0$ respectively. In the expansion~\eqref{eq:rho2}, the coefficients ${a_n}$ are not yet the physical deformation parameters, denoted in the following by ${q_n}$. It has been proved e.g. in \cite{schmitt2017} that the transformation between original ${a_n}$ amplitudes in the Fourier series ~\eqref{eq:rho2} and the physical deformation parameters ${q_n}$ can be of the following form: 
\begin{equation}\label{eq:q_to_a}
    \begin{split}
        q_2 &= a^0_2 / a_2 - a_2 / a^0_2, \\
        q_3 &= a_3, \\
        q_4 &= a_4 + \sqrt{(q_2 / 9)^2 + (a^0_4)^2}, \\
        q_5 &= a_5 - (q_2-2) \cdot a_3 / 10, \\
        q_6 &= a_6 - \sqrt{(q_2 / 100)^2  + (a^0_6)^2},
    \end{split}
\end{equation}
where parameters $a^0_2$, $a^0_4$, $a^0_6$ describe the spherical nuclear shape with radius $R_0$. 
In order to discuss the influence of non-axial shapes, one can easily modify our shape parametrization by multiplying the right-hand side of Eq.~\eqref{eq:rho2} by a function $f_{\eta}(\varphi)$
\begin{equation}
    f_{\eta}(\varphi)=\frac{1 - \eta^2}{1+\eta^2 + 2\eta\cos{\varphi}},
\end{equation}
chosen in such a way that any cross-section of the nuclear drop~\eqref{eq:rho2} perpendicular to the $z$-axis is an ellipse of half axes $a$ and $b$ while $\eta \equiv \frac{b - a}{b + a}$. 
The geometry of a non-axial, prolate nuclear shape is presented schematically in Fig.~\ref{f01}. The variety of the shape configurations is presented in Fig.~\ref{f02}.

The most relevant for fission process $\{q_2,q_3,q_4\}$ deformation parameters describe the nuclear elongation along $z$-axis, mass (volume) asymmetry of the left and right fragment, and the neck shape, respectively. It should be noted that the results presented in Refs.~\cite{schmitt2017, pomorski2021} reveal that the set of these three collective deformations $\mathbf{q}=\{q_2,q_3,q_4\}$ is sufficient to describe the behavior of the fissioning system close to its scission point within a reasonable energetical uncertainty of less than 1 MeV. Therefore, the higher order deformations, $q_5$ and $q_6$, which mainly modify the shapes of fission fragments in an insignificant way, are neglected at the current stage of our investigations. 

A similar argumentation applies the non-axiality degree of freedom, which is known to impact the PES of, in particular, actinide nuclei in the vicinity of the fission barrier, e.g. by reducing its height within 0.5-1 MeV. Thus the above property of PES in actinides allows us, at first approximation, to neglect the influence of the non-axial deformation $\eta$.

\subsection{Potential energy surface}

Setting the geometry of the nuclear surface, we come to the problem of defining the PES, which is a crucial factor determining the evolution of the fissile system.

From among a wide range of known approaches able to produce the potential energy function depending on the surface shape, we have decided to use a well-known macroscopic-microscopic model. Then, the total energy of a nucleus, $V(\mathbf{q})$, can be composed of the leading macroscopic term $E_{macr}$, evaluated in terms of a liquid-drop type approach, here the Lublin-Strasbourg Drop (LSD) \cite{pomorski2003}, while the microscopic interaction energy $E_{micr}$, playing the role of the energy correction on top of the dominating smooth liquid-drop term, is strictly related to the specific single-particle structure of a given nucleus
\begin{equation}\label{eq:Ecoll}
    V = E_{macr} + E_{micr}.
\end{equation}
The deformation-dependent LSD smooth energy contribution in~\eqref{eq:Ecoll} is written as
\begin{eqnarray}\label{eq:Elsd}
    E_{LSD}&=&b_{vol}(1-k_{vol}I^2)A-\nonumber\\
    &&b_{surf}(1-k_{surf}I^2)A^{2/3}B_{surf}(\mathbf{q})\nonumber\\
    &&-b_{cur}(1-k_{cur}I^2)A^{1/3}B_{cur}(\mathbf{q})\\
    &&-\frac{3}{5}e^2\frac{Z^2}{r_0^{ch}A^{1/3}}B_{Coul}(\mathbf{q})+C_4\frac{Z^2}{A}\nonumber\\
    &&-10\,exp(-4.2|I|), \nonumber
\end{eqnarray}
where $I = \tfrac{N - Z}{A}$ is the so-called {\it reduced isospin} whereas $B_{surf}$, $ B_{cur}$, $B_{Coul}$ introduce the deformation dependence to the surface, curvature, and Coulomb terms,  respectively.
The last deformation-independent term is what we usually call the congruence energy and is taken from the estimates of Myers and Swiatecki~\cite{myers1966}. All parameters of the LSD formula originally found in Ref.~\cite{pomorski2003} are also rewritten below:
\begin{eqnarray*}
&& b_{vol} = 15.4920 \ {\rm MeV}, \quad k_{vol} = 1.8601,\\
&& b_{surf} = 16.9707 \ {\rm MeV}, \  k_{surf} = 2.2038, \\
&& b_{cur} =  3.8602 \ {\rm MeV},\ \quad k_{cur} =-2.3764, \\
&& C_{4} = 0.9181 \ {\rm MeV}, \qquad r_0 = 1.21725 \ {\rm fm}.
\end{eqnarray*}

Please notice that this simple formula has been proven to reproduce the masses of over 3000 isotopes and over 80 fission barriers in actinides and super-heavy nuclei with reasonable accuracy.

In turn, the microscopic part in Eq.~\eqref{eq:Ecoll} is customarily decomposed into two energy components responsible for the shell, $E_{shell}$, and pairing interaction (superfluidity), $E_{pair}$, effects calculated within the Bardeen-Cooper-Schrieffer model proposed in \cite{bardeen1958}. 
The shell correction, $E_{shell}$ is, by definition, obtained by subtracting the mean energy $\tilde{E}$ arisen due to smoothing out the nucleon mean-field spectrum up to the levels from the energy continuum from the sum of the all occupied single-particle energies $e_k$ 
(see, e.g. \cite{dobrowolski2016})

\begin{equation}\label{eq:Eshell}
    E_{shell} = \sum_k {e_k} - \tilde{E}.
\end{equation}
In \eqref{eq:Eshell}, the averaged energy $\tilde{E}$ is estimated through the Strutinsky method~\cite{strutinsky1967, strutinsky1968} by smearing out the discrete spectrum
with a correction polynomial of the $6^{th}$ order. Finally, the pairing energy correction is determined in a similar way as done in \eqref{eq:Eshell}, but the resulting BCS energy is, in addition, reduced by the so-called average pairing-energy term, $\tilde{E}_{pair}$, which is not accounted in the smooth liquid-drop contribution \eqref{eq:Elsd}, as done in Ref.~\cite{nerlo-pomorska2012}
\begin{equation}
    E_{pair} = E_{\text{BCS}} - \sum_k{e_k} - \tilde{E}_{\text{pair}}.
\end{equation}
Single-particle spectra for protons and neutrons of here discussed actinide nuclei are eigenvalues of the folded-Yukawa mean-field Hamiltonian diagonalized numerically as described in Ref.~\cite{dobrowolski2016}.

\subsection{Nuclear shape evolution}

As mentioned, to describe the fission dynamics of selected actinide nuclei, we use a quasi-classical stochastic model, widely presented in Ref.~\cite{abe1986}. In this approach, a compound, excited, and in a general rotating nucleus is represented in the form of a superfluid incompressible drop~\cite{brack1972} with well-defined deformed surface whose time evolution is governed by the set of coupled Langevin equations as functions of collective deformation variables $\{q_i(t)\}$ and  the corresponding canonically coupled momenta $\{p_i(t)\}$, written as
%
%\begin{widetext}
    \begin{equation}\label{eq:Langevin}
        \left \{
            \begin{aligned}
                \dfrac{dq_i}{dt} &= \sum_{j} \left [ \mathcal{M}^{-1} \right ]_{ij} p_j, \\\
                \dfrac{dp_i}{dt} &= - \left[ \dfrac{1}{2} \sum_{jk}\frac{\partial \left[\mathcal{M}^{-1} \right ]_{jk}}{\partial q_i} p_j p_k + \dfrac{\partial F}{\partial q_i} \right. \\
                & \left. \quad\qquad + \sum_{jk} \gamma_{ij} \left [\mathcal{M}^{-1} \right]_{jk}p_k \right]   + \mathcal{R}_i,
            \end{aligned}
            \right.
    \end{equation}
%\end{widetext}
%
where $\mathcal{M}_{ij}$ and $\gamma_{ij}$ are tensors corresponding to mass (inertia) and friction, respectively, while $F$ is the Helmholtz free energy potential of the compound fissile system
\begin{equation}\label{eq:Helmholtz}
    F(\textbf{q}, T) = V(\textbf{q}) - a(\textbf{q}) T^2.
\end{equation}

In the above, $a(\textbf{q})$ is deformation-dependent energy level density, defined according to the prescription ~\cite{nerlo-pomorska2006}, and $T$ is the temperature of the system, which is related to the excitation energy $E^*$ through the relation:

\begin{equation} \label{eq:T}
    E^* = a(\mathbf{q})\,T^2.
\end{equation}
It is assumed that the excitation energy in our work, $E^*_{0}$, at an initial time $t=0$ is the difference of the excitation energy $E_{init}$ relative to the ground state and the height of the fission barrier $V_B$.

The last term, $\mathcal{R}_i$, of the second equation of the equation system~\eqref{eq:Langevin} corresponds to the $i^{th}$ component of the Langevin random force, which by definition writes
\begin{equation} \label{eq:Random force}
    \mathcal{R}_i = \sum_{j} g_{ij} \Xi_j(t),
\end{equation}
where $\Xi(t)$ is a time-dependent stochastic function given as $\Xi_j(t) =\nicefrac{\xi_j}{\sqrt{t}}$ with the following properties: $\langle \xi_k \rangle = 0$, $\langle \xi_k \rangle ^2 = 2$. The amplitudes $g_{ij}$ can be deduced from the fluctuation-dissipation theorem, known~\cite{kubo1966, abe1986} as the Einstein relation enabling for calculating the diffusion tensor
\begin{equation}\label{eq:Einstein}
    \mathbf{\mathcal{D}}_{ij} \equiv \sum_k g_{ik} g_{jk} = \gamma_{ij} \cdot T
\end{equation}
with $\gamma_{ij}$ being the friction tensor. The collective inertia used in Eq.~\eqref{eq:Langevin} is calculated within the incompressible irrotational flow approach using the Werner-Wheeler approximation~\cite{davies1976}  
\begin{eqnarray}\label{eq:M_tens}
\mathcal{M}_{ij}(\textbf{q})= \pi\rho_m \!\!\!\int\limits_{z_{min}}^{z_{max}}\!\!\!\! dz\rho^2_s(z, q)
\,\bigg[ A_i\,A_j\! +\! \frac{1}{8}\rho^2_s(z,q)A_i' A_j' \bigg], 
\end{eqnarray}
where $\rho_m = M_0 / (\frac{4}{3} R_0^3)$ is the average uniform nuclear density with $M_0= 0.0113 A^{\nicefrac{5}{3}}$ $[1/{\mathrm fm^3}]$, and the coefficients $A_i$ having the form:
\begin{equation}\label{eq:Ai}
A_i = \frac{1}{\rho^2_s(z,\mathbf{q})}\frac{\partial}{\partial q_i}\int\limits_{z}^{z_ {max}}
\rho^2_s (z',\mathbf{q}) dz'.
\end{equation}
Since our discussion is limited to the low energy fission of compound nuclei, the usage of the hydrodynamic macroscopic approach to describe the friction tensor $\gamma_{ij}$ is convenient. Then the one-body dissipation friction-tensor component may be expressed through the so-called "wall" formula~\cite{bartel2019} as 
\begin{equation}\label{eq:wall}
    \gamma^{wall}_{ij} = \frac{\rho_m}{2} \bar{v}\int\limits_{z_{min}}^{z_{max}}{\frac{\frac{\partial \rho^2_s}{\partial q_i} \frac {\partial \rho^2_s}{\partial q_j}}{\sqrt{4 \rho^2_s + \left( \frac {\partial \rho^2_s}{\partial z} \right)^2}}}dz.
\end{equation}
In the above, $\bar{v}$ is the average intrinsic velocity of nucleons, the value of which can be evaluated within the Fermi gas model to be equal $\bar{v}=\frac{3}{4}v_{F}$, where $v_{F}$ is the nucleon velocity at the Fermi level. 

\subsection{Temperature effects}

As it follows from Eq.~\eqref{eq:Helmholtz}, the system of Langevin equations~\eqref{eq:Langevin} has an additional hidden parameter, namely the temperature $T$, which is necessary to obtain a correct solution in case of substantially excited systems. The information about the change of $T$ during nuclear shape evolution from the initial to final (scission) state can be extracted from the energy conservation low, to be fulfilled at each time along a given trajectory and has the following form:
\begin{equation}\label{eq:E_total}
E_{total} = \frac{1}{2} \sum_{jk} \left [ \mathcal{M}^{-1} \right ]_{jk} p_j p_k + V(\mathbf{q},T) + E^*,
\end{equation}
where the collective macroscopic-microscopic potential $V$ contains a temperature dependent microscopic quantum correction, described in Ref.~\cite{nerlo-pomorska2006} by the relation
\begin{equation}\label{eq:E_mic_T}
    E_{mic}(\mathbf{q}, T) = \frac{E_{mic}(\mathbf{q}, T = 0)}{1 + e^{\frac{1.5 - T}{0.3}}}.
\end{equation}
Hence, with increasing excitation, the viscosity of the nuclear liquid must change and finally lose its superfluid properties. As shown in~\cite{ivanyuk1996}, the temperature dependence of the friction tensor must have a similar dependence as the above shell correction, where the temperature-dependent coefficient writes
\begin{equation}
    \gamma^T_{ij}(\mathbf{q},T) = \frac{0.7}{1 + e^{\frac{T_{\gamma} - T}{a_{\gamma}}}}\, \gamma^{wall}_{ij}(\mathbf{q},T=0)
\end{equation}
with the constants $T_{\gamma}=0.7$ MeV and $a_{\gamma}=0.25$ MeV providing good description of the dissipative properties of the diffusion tensor $D_{ij}$ of Eq.~\eqref{eq:Einstein}. The introduced temperature-dependent factor significantly changes the friction when $T$ tends to zero, e.g. in spontaneous fission.
As known, classical Brownian motions vanish when the system's temperature tends to zero. Thus the diffusion tensor $D_{ij}$, which fixes the magnitude of the random Langevin force, should vanish, too and therefore the statistical nature of the fission processes will be violated. On the other hand, quantum-mechanical considerations bring us to the conclusion that even in temperature being close to zero, i.e. for very low excitation energy, the zero-point motion of nucleons can cause fission.

To simulate these quantum effects in the semi-classical Langevin description, one can replace the temperature $T$ with an effective temperature $T^*$ in ~\eqref{eq:Einstein}, as proposed in Ref.~\cite{hofmann1998}
\begin{equation}\label{eq:T*} 
    T^* = E_0 \coth \frac{E_0}{T},
\end{equation}
where $E_0 = \tfrac{\hbar \omega_0}{2}$ corresponds to the zero-point collective oscillation energy of the nucleus in the vicinity of its ground state, which typically varies between $0.5 - 2$ MeV. Under this assumption one obtains from Eq.~\eqref{eq:Einstein} a more realistic description of the friction in low energy fission, responsible for the energy exchange between single-particle and collective degrees of freedom.

The set of Langevin equations~(\ref{eq:Langevin}) is solved by the discretization method in which the corresponding differential quotients are applied instead of the time derivatives on the left-hand sides of both equations. The finite time step for the numerical solution of their discretized forms is taken as $0.01 \tau$ of the characteristic relaxation time $\tau \equiv \dfrac{2 \mathcal{M}}{\gamma} \dfrac {\hbar} {\text{MeV}}$.

\subsection{Initial and trajectory-terminating conditions}

Having described the essential components of the model, we can proceed to a crucial point of this work, namely, defining the set of boundary conditions for the differential Langevin equations. For this purpose, first, one should define a region in the domain of collective variables $\mathbf{q}$ in which the shape evolution of a nucleus is performed. Some detailed studies have shown that for actinide nuclei, it is necessary to consider the following collective three-dimensional deformation space
\begin{align}\label{grid}
    q_2 &= \left[ \ \ \ 0 \quad \ (0.05) \ 2.35 \right]\nonumber \\
    q_3 &= \left[ -0.21\   (0.03) \ 0.21 \right]\\
    q_4 &= \left[ -0.21\   (0.03) \ 0.21 \right] \nonumber
\end{align}
which comprises a vicinity of the ground state, all relevant for fission process saddle points and isomeric minima ending within the configurations, where the nucleus is already split into two fragments, or alternatively, the width of the neck of a compound nucleus is sufficiently small (around $0.2\,R_0$) to observe the fission.
In the nodes of such lattice, we have calculated the previously introduced values of the collective potential, inertia, and friction tensors. To determine the values between the lattice nodes, we use the so-called Gauss-Hermite approximation method proposed in Ref.~\cite{pomorski2006}, which determines the demanded values on, in general, N-dimensional mesh with very satisfactory accuracy. 

At the end of the paragraph, let us mention the behavior of a trajectory when a variable $q_i$, being a part of the parametric definition of that, reaches its extreme (border) value given in (\ref{grid}). This may happen, for example, when the entry point in a given isotope is located relatively close to the grid boundaries and, therefore, after a couple of time steps, can easily reach the border. Technically, such a trajectory does not lead to fission and, strictly speaking, should be removed from our consideration. In such a case, some conditions for resuming such a trajectory may be helpful. A reasonable possibility to solve this problem may be to change the sign of the momentum component conjugated to this coordinate, allowing it to turn around and continue its evolution.

With the coordinate $q_2$, the situation is slightly exceptional. After reaching its maximum value $q^{max}_2 = 2.35$, the system is elongated more than two times than in its ground state. Suppose that for such a deformation, the decisive criterion for qualifying a given trajectory as the one which leads to fission is still not fulfilled. In that case, such a trajectory has no physical meaning, and its further evolution is meaningless. Hence, this kind of basic condition must be imposed in most further calculations, mainly if a symmetric and highly elongated fission channel is intensely populated.

\section{Effect of boundary conditions on model results}

Nevertheless, previously mentioned boundary conditions usually need to be improved to determine stochastic trajectories effectively. We initiate the evolution of a Langevin trajectory by choosing its initial deformation point on the PES. In general, when using the mentioned formalism, such a point is assumed to be the ground state of the compound system, as it is done in the works of Abe~\cite{abe1986, abe1996}. According to the theory of our stochastic framework, the compound system has to initially stay close to the ground state to be able to say about its evolution towards different decays. 
In fission, the available energy excess has, at least, to be enough for the system to overcome the barriers standing in the way of the fissioning nucleus. Thus, the set of initial configurations drawn at the beginning of each trajectory is likely located in a specific area in the vicinity of the outer saddle point, through which the system must pass in the direction of fission.

\subsection{Modifications of initial conditions}

In order to prove this assumption, we calculate the fragment mass distributions for the thermal neutron-induced fission of $\rm ^{235}U$ nucleus with two different starting points. The first option is to initiate the trajectory from the ground state. Using the initial conditions mentioned in the previous paragraph, the second one starts from the outer saddle point. The initial conjugated momenta are put in both cases to be equal to zero.

We assume that a nucleus undergoes fission and the determination of the corresponding trajectory is terminated if the neck radius in the thinnest point reaches $r_{neck}\approx 0.3\,R_0=2.0$ fm (see, e.g., \cite{mazurek2017, liu2019, liu2021}. Of course, such a criterion is chosen, to some extent, arbitrarily and can be modified by introducing a dependence of $r_{neck}$ on the scission deformation (elongation) or temperature. Nevertheless, in this study, no such dependence is assumed. 
\begin{figure}
    \centering
    \includegraphics[width=.975\linewidth]{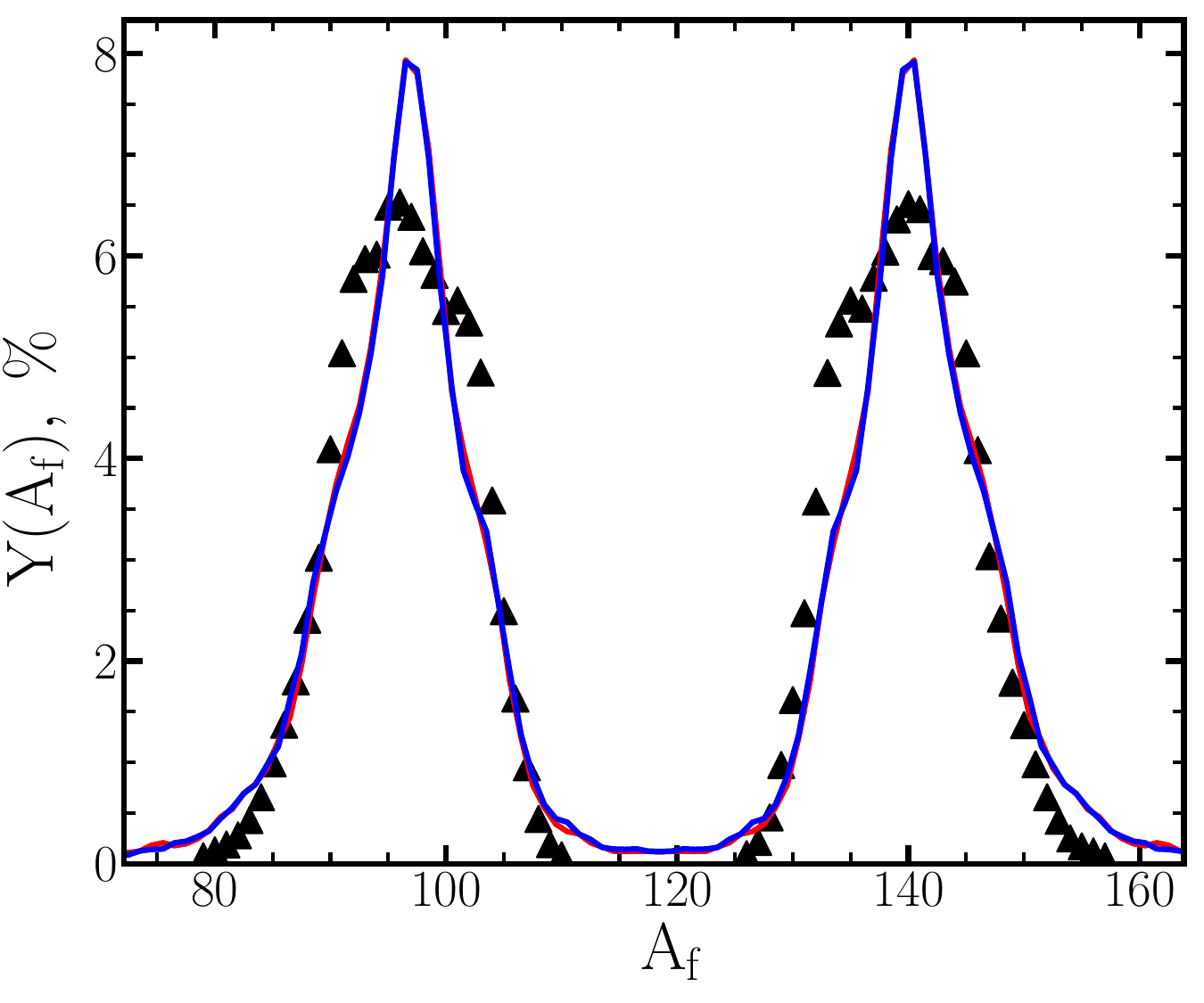}
    \vspace{-10pt}
    \caption[]{Primary FMDs for thermal neutron induced fission of $\rm ^{235}U$ initiated from the ground state (red) and the second saddle (blue) whereas black triangles correspond to values adapted from experimental data \cite{simon1990}. Here $Y(A_f)$ denotes the yield for the corresponding fragment mass.}\label{f03}
\end{figure}
As seen in Figure~\ref{f03}, the fragment mass distributions for both these cases are nearly identical.
 The total number of trajectories used here to generate serious statistics is significant and equal to $10^5$. However, only $1$ per $100$ initiated in the ground state trajectories overcome the barrier and efficiently evolved to fission. In contrast, the others are stuck in the potential energy well for a long time. If the calculation starts in the direct surrounding of the saddle point, the number of "imprisoned" trajectories is lower practically by some order of magnitude.
However, introducing a few additional constraints can still improve the ratio of "passed" to "imprisoned" trajectories. We, therefore, use the method according to the ideas proposed in the Refs~\cite{krappe_pomorski, kostryukov_2021}, where the manner of generating $\mathbf{q}^0$ initial coordinates to be used in the first time step was proposed. This method consists of the following procedure: using the normal distribution $\xi_{norm}$ with $\mu=0$ and $\sigma=\frac{1}{2}\sqrt{\nicefrac{E_0}{\frac{\partial^2 V}{\partial q_i^2}}})$ we fix the set of coordinates $\mathbf{q}^0$,  which then has to satisfy the following condition
% $V(\mathbf{q}^{start}) - V(\mathbf{q}^0) > 0$.

\begin{equation}\label{eq:orig_cond}
    \left \{
        \begin{aligned}
            &q^0_2 \ge q^{start}_2\\
            &\frac{1}{2} \sum_{ij} \left [\mathcal{M}^{-1} \right ]_{ij} p^0_i p^0_j \equiv V(\mathbf{q}^{start}) \, - \\
            & \qquad \qquad - V(\mathbf{q}^0) - E_0  \ge 0.
        \end{aligned}
    \right.
\end{equation}
where $\mathbf{q}^{start}$ is an actual starting point of a trajectory, and $E_0$ describes a contribution of the zero-point vibration energy at that point.

\begin{figure*}
    \centering
    \includegraphics[width=0.95\textwidth]{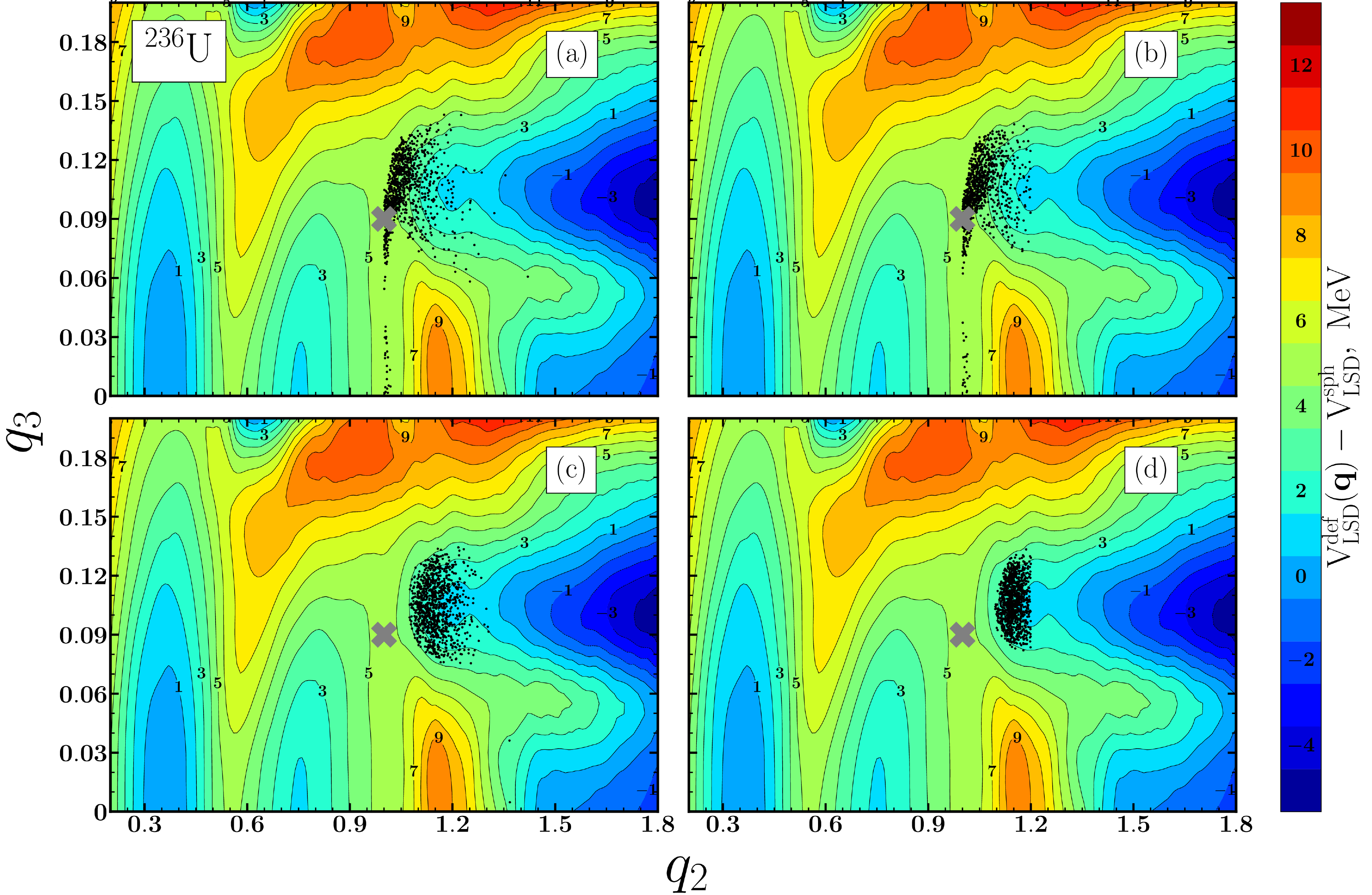}
    \caption[]{Samples of starting-point distributions on PES of $\rm ^{236}U$. Panel (a) - without limit control and subtraction of $E_0$, (b) - without limit control and inclusion of $E_0$, (c) - without control and inclusion of $E_0$ and (d) - with limit control and $E_0$ included, the gray cross gives the location of the second saddle.}\label{f04}
\end{figure*}

The question arises whether the space of $\mathbf{q}^{0}$ points should be restricted to a certain volume around the point $\{\mathbf{q}^{start}\}$. In terms of the condition (\ref{eq:orig_cond}), such a problem may occur when the PES is sufficiently flat around this point, allowing the initial configuration to exceed the borders of the fixed grid, see Fig.~\ref{f04}(a,c). To avoid this, we can somewhat arbitrarily restrict the deformation space $\mathbf{q}^{0}$ to the following limits:
\begin{equation}\label{eq:q1_lim}
    \begin{aligned} 
        q_2 &= \left[ q^{start}_2; \ q^{start}_2 + 0.2 \right]\\     
        q_3 &= \left[ q^{start}_3 \ - \ 0.09; \ q^{start}_3 + 0.09 \right]\\
        q_4 &= \left[ q^{start}_4 \ - \ 0.09; \ q^{start}_4 + 0.09 \right]
    \end{aligned}
\end{equation}
In Fig.~\ref{f04}, we see four PES for $\rm ^{236}U$, where the coordinates $(\mathbf{q}^0$ are distributed without (a,b) and with (c,d) including the zero-point vibration energy in (\ref{eq:orig_cond}) and with and without limitations of the initial coordinate region (b,d). The figure in the last two panels reveals the lack of sensitivity to these limitations. In this case, the ratio of traversed to not traversed trajectories in case (d) places within the interval $1-1.5$, noticeably reducing the computation time.

\begin{figure}
    \centering
    \includegraphics[width=\linewidth]{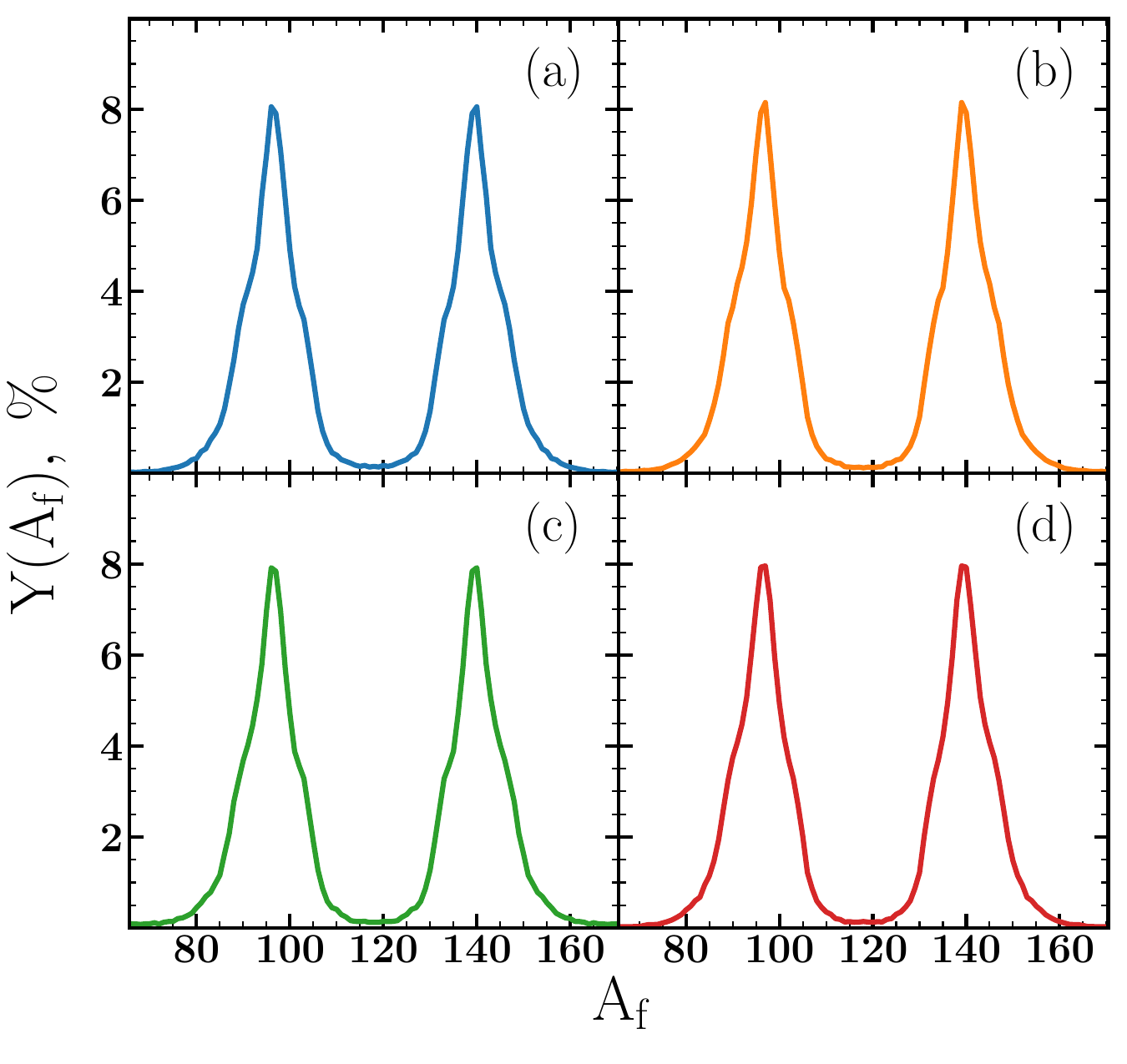}
    \vspace{-12pt}
    \caption[]{Primary FMDs for starting-point distributions, where symbol $(i)$ corresponds to analogous cases from Fig.~\ref{f04}.}\label{f05}
\end{figure}

\subsection{Fissioning trajectories}\label{fiss_traj}

Having determined the criteria for fixing the starting point for a given Langevin trajectory which is assumed to lead to fission, let us turn to the problem of assessing whether, at a given time, the trajectory describes a fission configuration or whether the nucleus is still compound. This task, as commonly known, is not trivial as, in reality, the division of a nucleus into fragments may significantly depend not on the neck width alone but also on a series of other quantities characterizing bulk and surface properties of both fragments, their shell structures, deformations, excitation energies, the relative collective velocity of fragments towards fission, neck curvature, etc. As we have explored, knowing the decisive criteria for suspending the evolution of a given trajectory due to the achievement of a neck braking configuration is even more crucial than choosing its starting point. Unfortunately, this problem is not unambiguously solvable at the moment. It requires the introduction of several additional phenomenological assumptions, which will only be tested by comparing the simulation results with empirical data. This definitively may reduce the transparency and universality of this approach.

Since the phenomenological criteria for the neck rapture are, to some extent, arbitrary and model dependent, we decide to test within this work the one which effectively would lead to a division of a nucleus into two fragments and depending only on the neck radius (width), $r_{neck}$, in case axial shapes are considered. Such a solution is widely used in several recent works, e.g., Refs.~\cite{abe1996, adeev2005, sierk2017, mazurek2017, liu2019, liu2021}. In our approach, the fission onto two fragments occurs when the neck radius, a value of which may vary between $1-2.5$ fm, is close to the effective radius of a single nucleon, denoted in the following by $r_n$ and is approximately equal to 1 fm. 
Suppose the neck radius $r_{neck}$ is too big at reaching the presumed elongation grid-border $q^{max}_2$. In that case, the trajectory has, in principle, should be excluded from consideration as a non-fissioning one.

%\subsection{Checking and limiting finite configurations}
In practical calculations on a finite deformation grid, such grid-border values are usually fixed slightly before geometrical scission, i.e., where the neck radius is strictly equal to zero. This is so because the accuracy of numerical determining of the PES and necessary transport quantities for already two separated, strongly elongated fragments are considerably lowered due to limitations of numerical routines used to develop the eigensolutions of the Yukawa-folded Hamiltonian and the liquid-drop deformation functions in highly elongated, necked nuclear shapes.

\begin{figure}
    \centering
    \includegraphics[width=0.9\linewidth]{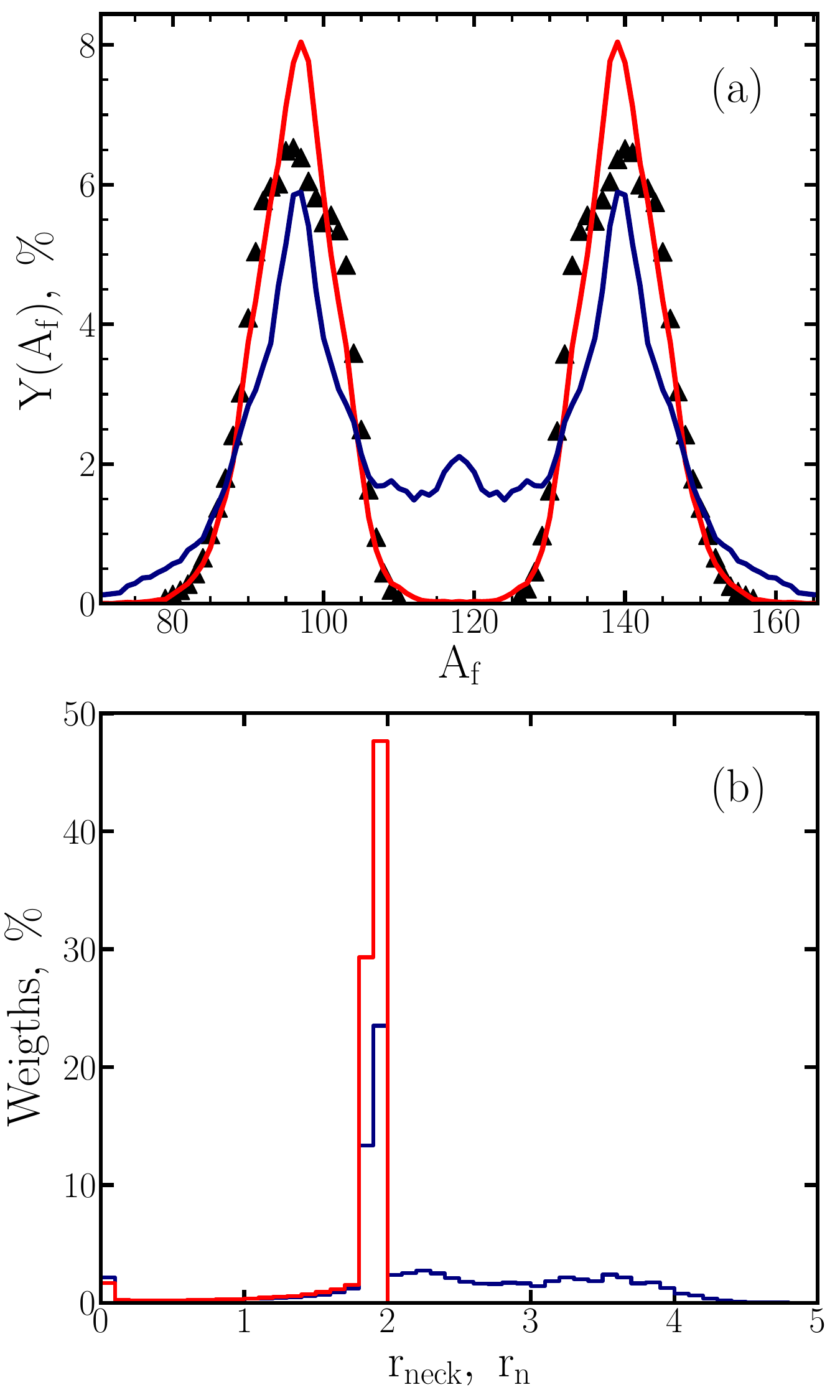}
    \vspace{-10pt}
    \caption[]{Primary FMD's (a) for thermal neutron induced fission of $\rm ^{235}U$ with obligatory usage of neck radius condition (red) and without (navy). The histogram (b) shows $r_{neck}$ distribution for both cases.}\label{f06}
\end{figure}

However, in specific test cases shown below, where the scission configurations for symmetric fission can be strongly elongated, we allow for a possibility that the trajectory is continued even though the presumed elongation limit, $q^{max}_{2}$, is slightly exceeded. At the same time, the condition for the neck radius is still not satisfied.
To show the contribution of such strongly elongated states to the final FMD, in addition to the previously obtained initial conditions~(\ref{eq:q1_lim}), we introduce the trajectory-termination conditions in two ways. First of them, for elongations $q_2<q^{max}_2$, a trajectory that satisfies only the neck-radius criterion $r_{neck} < r^{stop}_{neck}$ is counted as a fissioning one. In case $q_2 > q^{max}_2$ and the neck radius is still greater than a fixed $r^{stop}_{neck}$ value, such a trajectory is then rejected. This scenario is depicted in Fig.~\ref{f06}(a) with the red line. In contrast, we consider the second way, where the neck radius condition is completely ignored, and the trajectory reaching the elongation limit $q^{max}_2$ describes the act of fission. Clearly, in the last scenario, the values of the neck radii in the fissile configurations distribute over different possible values ranging from $r_n$ to even more than $4r_n$ with a clear peak around $2r_n$, as presented in Fig.~\ref{f06}(b) with the navy blue line.

\begin{figure}
    \centering
    \includegraphics[width=.95\linewidth]{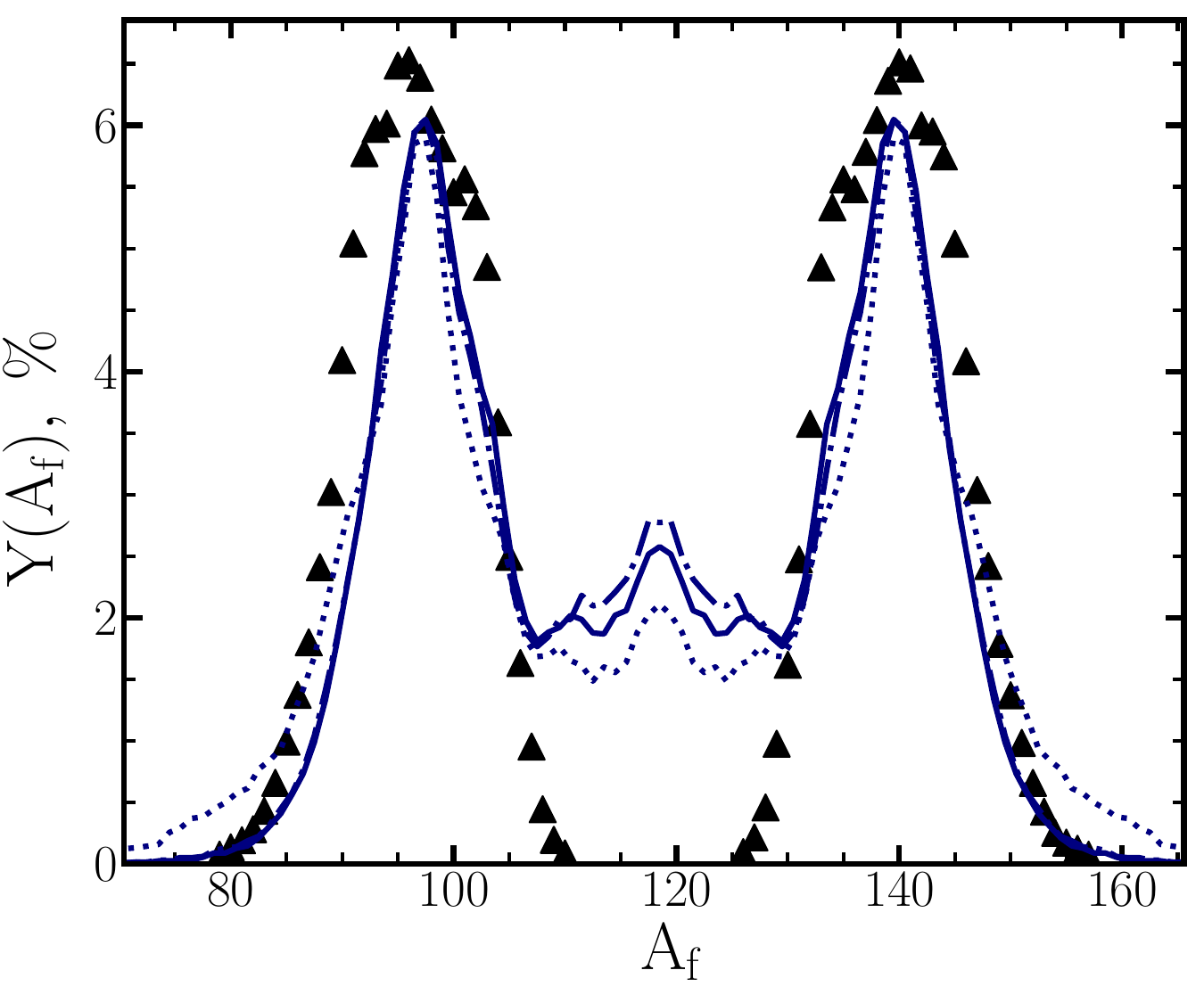}
    \vspace{-10pt}
    \caption[]{Primary FMDs for thermal neutron induced fission of $\rm ^{235}U$ with non-obligatory neck radius condition usage at limit value $q^{max}_2 = 2.35$ (dotted line), $q^{max}_2 = 2.5$ (dash-dotted line) and $q^{max}_2 = 2.9$ (solid line).}\label{f07}
\end{figure}
As shown in Fig. ~\ref{f06}(a), neglecting the "neck-radius condition" results in a significant contribution of both near-symmetric and extremely asymmetric channels, which are not observed in the experimental distribution. To explain this, let us return to Fig.~\ref{f02}, where one can see that at elongation $q_2 = q^{max}_2 = 2.35$, in light of the neck-radius condition, and some configurations cannot be qualified as being very close to splitting. If one considers the achievement of this elongation limit as the only decisive condition for fission, there appears a danger of obtaining unrealistic FMD. As also seen in Fig.~\ref{f02}, extending this value even to $q^{max}_2 \approx 2.9$, at which the accuracy of determining necessary input quantities is getting increasingly questionable, we are still facing nuclear shapes with significant neck widths of approximately $0.5R_0$. Figure \ref{f07} displays the increase of the near-symmetric fission yields of the FMD with a gradual shifting of the $q^{max}_2$ value from $2.35$ to $2.9$. The above is an effect of the undesired property of our Fourier shape parametrization, which, particularly for large nuclear elongations, cannot produce well-separated, symmetric fragments.

One then deduces that the conditions for nuclear scission applied to our Langevin framework, which mainly determines the quality of reproduction of FMD, have to be searched according to the following rules: first, by considering pure geometrical criteria for the neck width, dependent only on the surface-parametrization properties and second, by verifying whereas, for such pre-selected deformation point, the accuracy of determining the macroscopic-microscopic quantities fit the acceptable limits. This also indicates that the choice of the optimal $r^{stop}_{neck}$ value may not, in general, be universal across the complete set of studied nuclei and needs, at least, to be validated when changing Z or N numbers by a couple of units.

\subsection{Searching for the optimal neck radius}

\begin{figure}
    \centering
    \includegraphics[width=\linewidth]{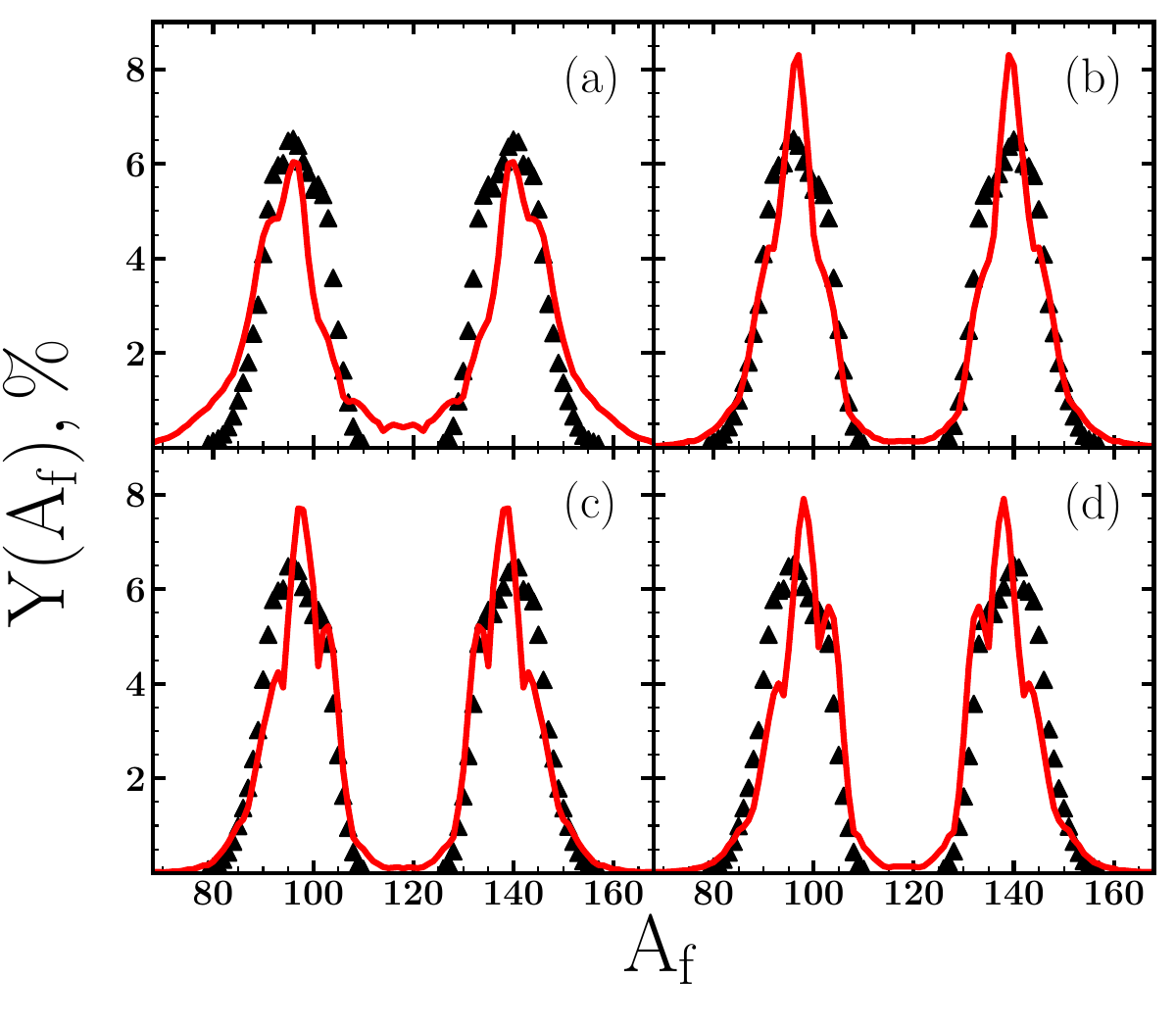}
    \vspace{-10pt}
    \caption[]{Primary FMD's for neutron-induced fission $\rm ^{235}U$ with variation of the neck radius $r_{neck}$ from 3$r_n$ (a), 2$r_n$ (b),  $r_n$ (c), 0 (d).}\label{f08}
\end{figure}

Now, after proving that the condition of the neck size is crucial, let us investigate its effect on the distributions of the fission mass fragments. For this purpose, we assume that the value of the limit radius $r^{stop}_{neck}$ at which a trajectory is stopped may vary from $3r_n$ to $0$ with a step of $r_n$. We set the initial points according to Fig.~\ref{f04}(d) and prescription of Eq.~(\ref{eq:q1_lim}) while the upper elongation limit $q^{max}_2=2.35$. As can be seen from Fig.~\ref{f08}, the resulting mass distributions change their form for different $r^{stop}_{neck}$ radii. With decreasing neck radius, the fragment mass distribution is getting slightly narrower, and the asymmetric peak shifts towards more and more symmetric yields. At the same time, its symmetric part is gradually vanishing, approaching the experimental value.

To understand the dominance of the asymmetric fission channel in this nucleus, let us notice at the PES presented in Fig.~\ref{f04} that the most like path from the starting configuration, set around the second saddle point at $(q_2,q_3)\approx (1.0,0.09)$, to the scission leads directly towards the asymmetric valley which is separated from the symmetric one by the edge of almost 3 MeV high, visible around $q_3\approx 0.06$. Moreover, since the excitation energy at the initial configuration is relatively low, the random force defined through Eqs.~\eqref{eq:Random force} and~\eqref{eq:Einstein} has very little chance to push the system over this edge.
It can also be seen that except for the extreme cases (a) and (d) with $r^{stop}_{neck}=3\,r_n$ and $r^{stop}_{neck}=0$, respectively, the overall features of other presented distributions are generally weakly affected, which may indicate that the main contributions to the final FMD come from $r^{stop}_{neck}=\{2r_n, \ r_n, \ 0 \}$.

\subsection{Stochastic character of neck-breaking }

\begin{figure*}
    \centering
    \includegraphics[width=0.9\linewidth]{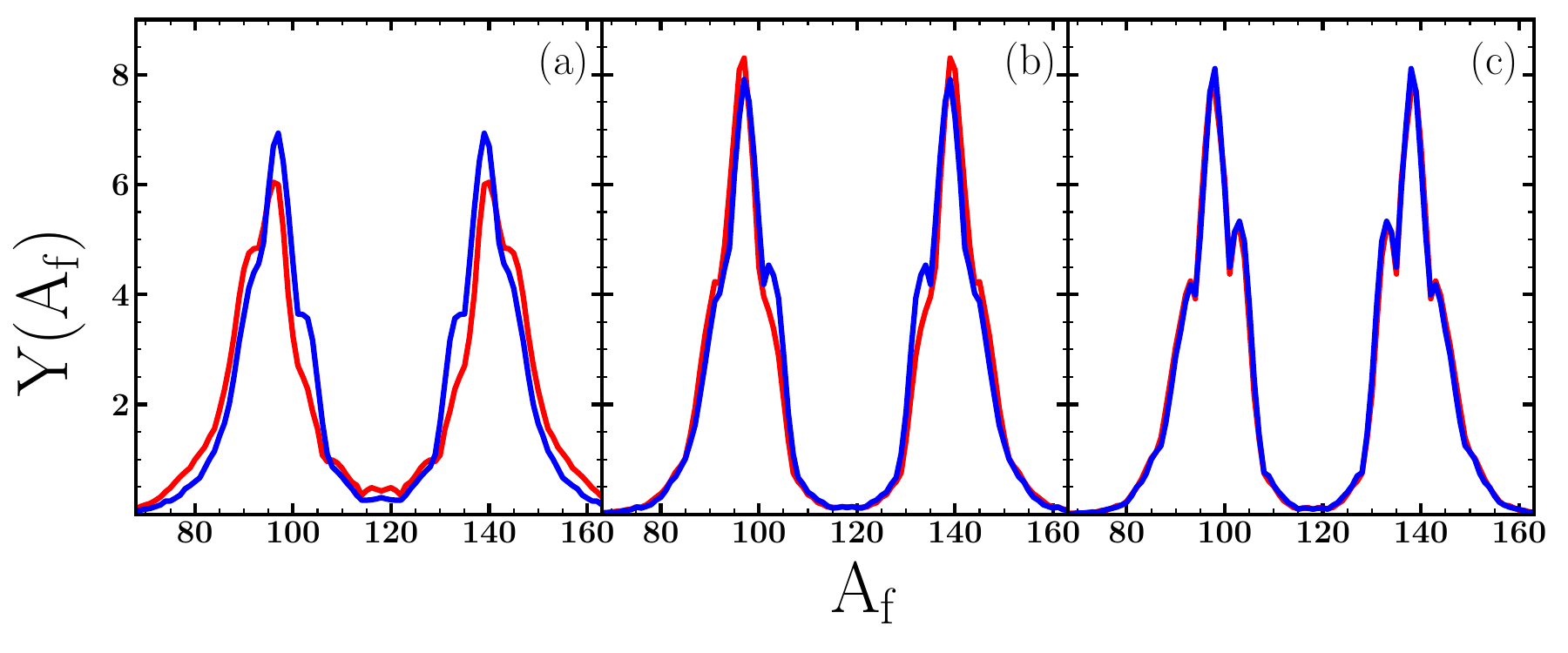}
    \vspace{-10pt}
    \caption[]{Primary FMD's for thermal neutron induced fission of $\rm ^{235}U$ calculated within random pick (blue) of $r^{stop}_{neck}$ defined on the following intervals [0,3$r_n$] (a), [0, 2$r_n$] (b) and  [0,$r_n$] (c), compared with analogous FMD's of Fig.~\ref{f08}(a)-(c).}\label{f09}
\end{figure*}

Taking into account the results shown in Fig.~\ref{f08}, one can ask whether the use of the strictly fixed value of the $r^{stop}_{neck}$ which governs the moment of splitting of a nucleus into fragments of different masses (charges) is not a severe simplification of the stochasticity of fission phenomenon. A simplistic realization of the idea of, to some extent, random value of the $r^{stop}_{neck}$ radius just before the neck-breaking is to draw at the beginning of each trajectory its value from a specific interval, say $[0,\alpha_r\,r_n]$, with a probability given through the uniform distribution. The fragment mass distributions shown in Fig.~\ref{f09} are calculated for the following three values of $\alpha_r=\{1,2,3\}$. The results are compared with the FMD obtained within analogous intervals shown in Fig.~\ref{f08}(a)-(c), respectively.

\vspace{12pt}
\begin{figure}
    \centering
    \includegraphics[width=\linewidth]{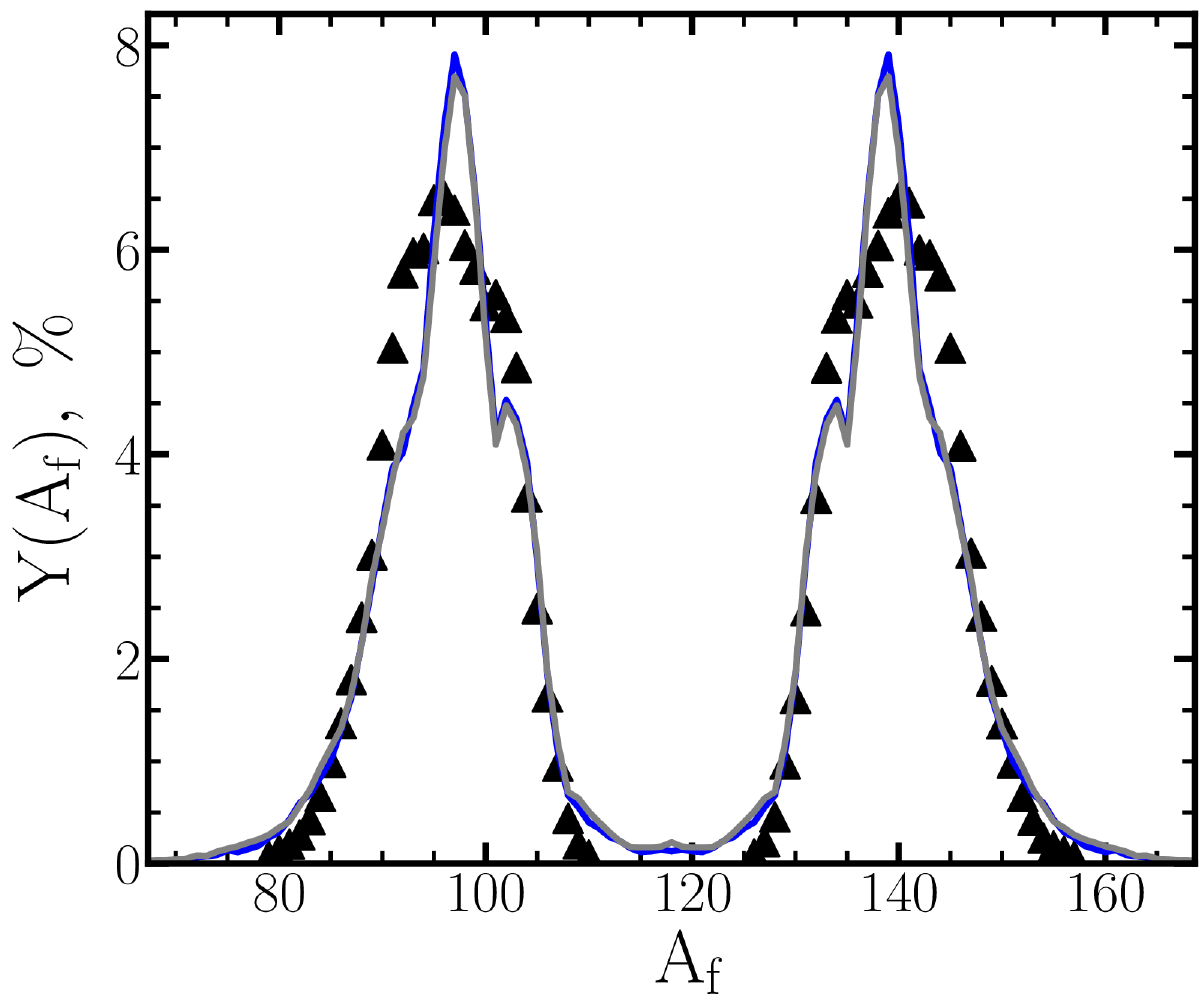}
    \caption[]{Comparison of primary FMD's for thermal-neutron induced fission $\rm ^{235}U$ calculated within Gaussian $P_{norm}(r_n,r_n)$ (gray) and random picking (blue) distributions imposed on $r^{stop}_{neck}$.}\label{f10}
\end{figure}

\vspace{-12pt}
The above concept may also be realized if instead of the uniform discrete distribution of $r^{stop}_{neck}$ one uses the continuous normal distribution peaked at $r_n$ with the dispersion $\sigma$ equal to $r_n$, denoted by $P_{norm}(r_n,r_n)$. These parameter values allow us to cover all the scission neck configurations previously considered in Fig.~\ref{f09}. If the drawn value of $r^{stop}_{neck}$ happens negative, its absolute value is taken. The resulting distributions seen in Fig.~\ref{f10} with comparison to the previous ones of Fig.~\ref{f09}(b) seem to be hardly distinguishable.

\subsection{Final conditions and excitation energy}

\begin{figure}
    \centering
    \includegraphics[width=\linewidth]{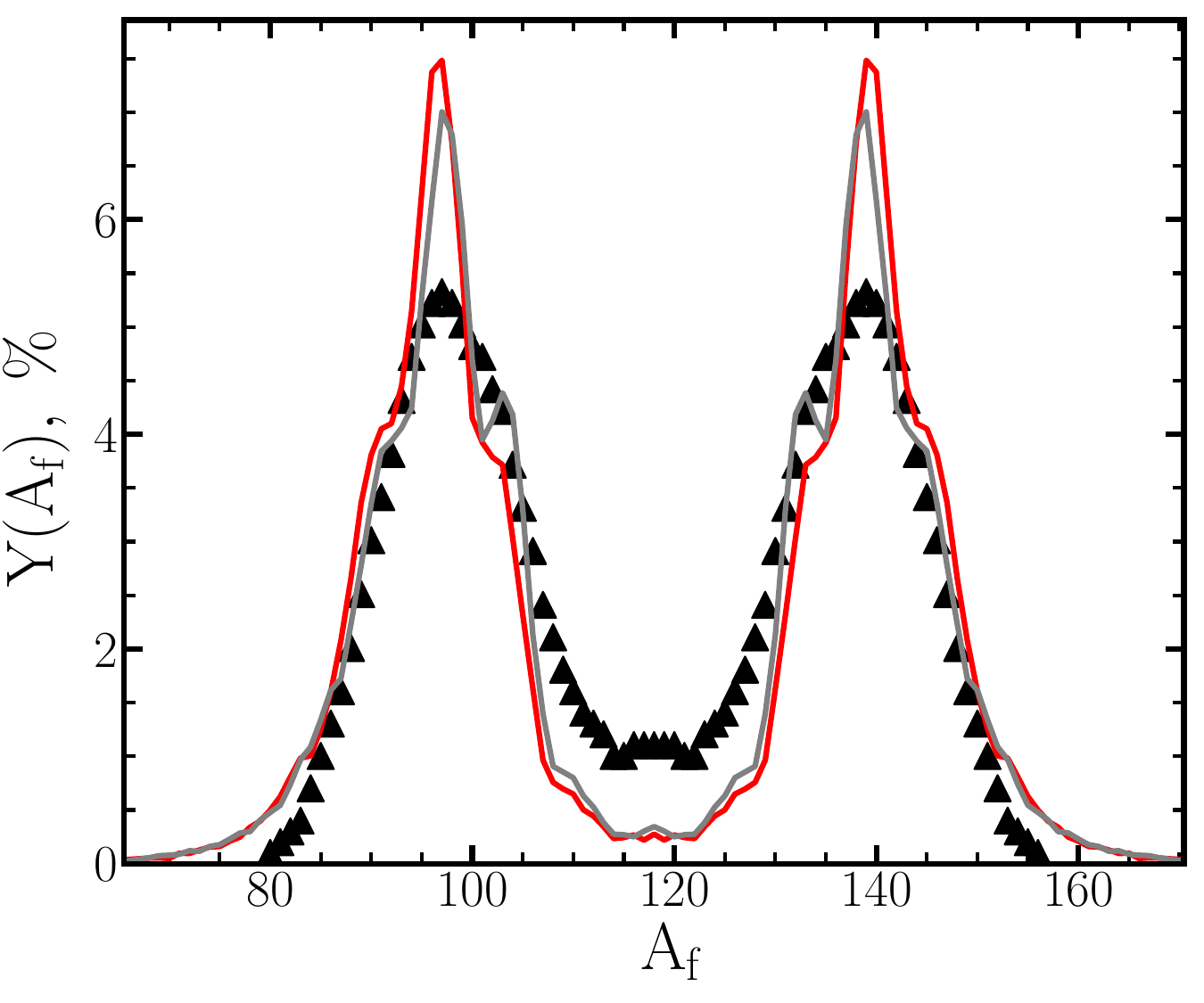}
    \caption[]{Comparison of primary FMD with experimental data~\cite{dyachenko1969} for 15 MeV neutron induced fission of $\rm ^{235}U$ (black triangles) with FMDs calculated within Gaussian 
    $P_{norm}(r_n,r_n)$ distribution (gray) and constant value $2r_n$ (red) of $r^{stop}_{neck}$.}\label{f11}
\end{figure}

As can be seen, the introduction of a more involved Gaussian distribution on $r^{stop}_{neck}$ thresholds does not qualitatively change the final fragment mass distributions for thermal neutron-induced fission of $\rm ^{235}U$. One can then apply a similar procedure to analyze the shapes of distributions for the systems of higher excitation energy. It is clear that at higher temperatures, the system, especially in the neck region, is less stable. Some local surface vibration provoked by thermal nucleon motion can lead to a more rapid neck rupture, even when its radius is much greater than $r_n$.
We then study the fast-neutron fission reaction where $E_n = 14.8$ MeV. This means the excitation energy $E^*$ exceeds the fission barrier $V_B$ by almost 15 MeV. The calculation is performed for the two variants of $r^{stop}_{neck}$ conditions, where first, $r^{stop}_{neck} = 2r_n$ (see further) and second, $r^{stop}_{neck}$ is randomly drawn with the probability given by the Gaussian distribution 
$P_{norm}(r_n,r_n)$. The results are shown in Fig.~\ref{f11}. It is seen that both theoretical estimates of FMDs have a serious discrepancy with the experimental data in the region of symmetric channels.
%It is seen that for both theoretical estimates of FMDs, a severe discrepancy with the experimental data in the region of symmetric channels is noticed.
%

\begin{figure}
    \centering
    \includegraphics[width=\linewidth]{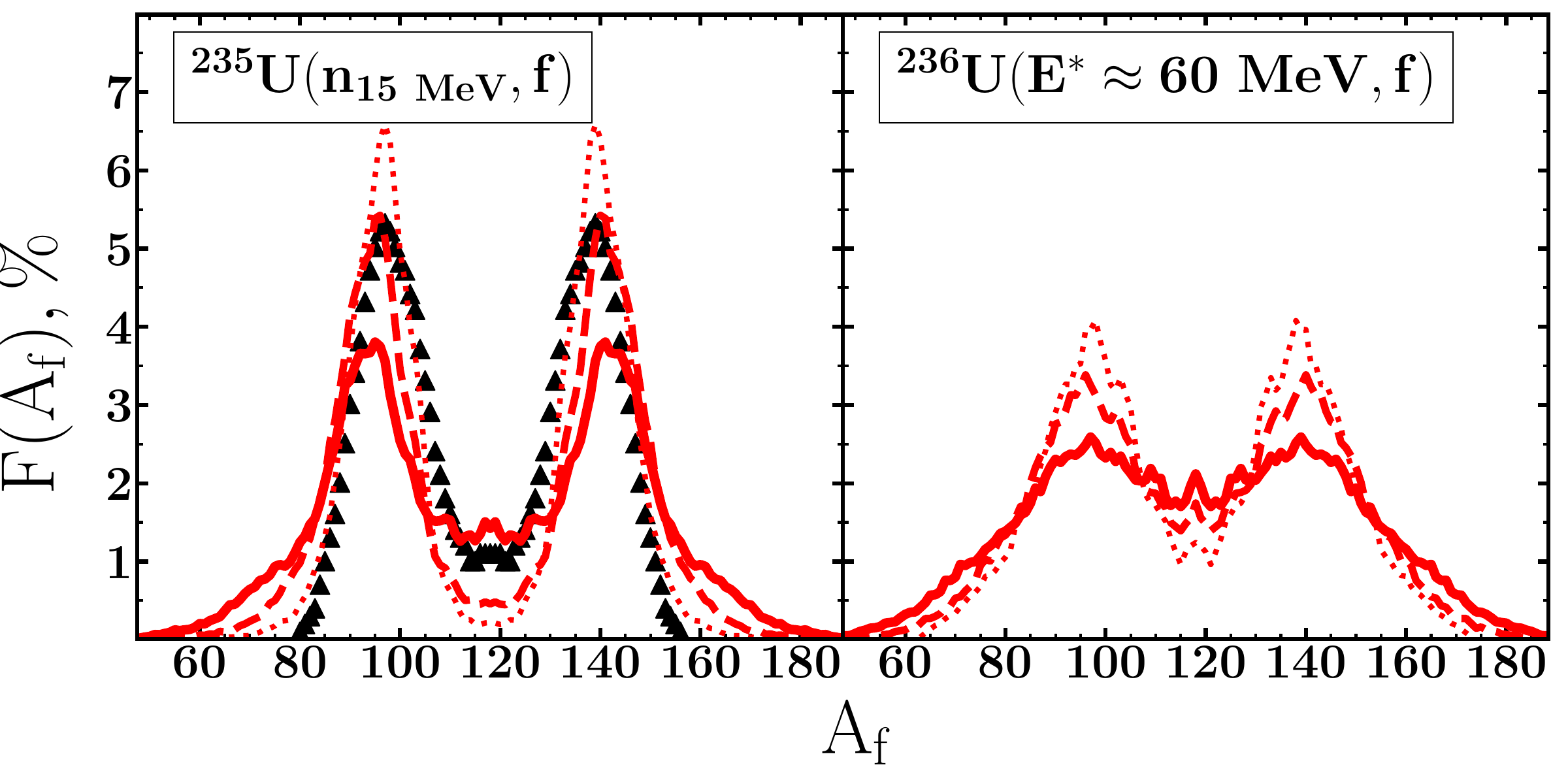}
    \vspace{-24pt}
    \caption[]{Comparison of the primary FMDs for 15 MeV neutron-induced fission $\rm ^{235}U$ (left) and $\rm ^{236}U$ with 55 MeV excitation above the barrier top (right), calculated for the value of $r^{stop}_{neck}$ varying from 2$r_n$ (dotted line), 3$r_n$ (dashed line), 4$r_n$ (solid line). Experimental data~\cite{dyachenko1969} are also presented to make the comparison clearer.}\label{f12}
\end{figure}

This example illustrates, slightly in contrast to the thermal-neutron induced fission depicted in Fig.~\ref{f10}, an increased sensitivity of the FMD on the conditions which define the end of a Langevin trajectory. A good illustration to that statement is shown in Fig.~\ref{f12}, where the neck-radius $r^{stop}_{neck}$ varies from 2 to 4 values of $r_n$ for the previously considered system with an excitation energy of about 15 MeV above the barrier (left panel). For comparison, we consider the uranium system with excitation energies already of 55 MeV in the right panel.
With the condition for neck radius $r^{stop}_{neck} = r_n$ the form of the FMD is practically the same as for $2r_n$ while at higher values, e.g. $r^{stop}_{neck} > 4 r_n$, it is difficult to talk about the occurrence of true neck.
One can see that the symmetric yields of the final FMD's become closer to the measured values when the neck radius is higher than in the case of the thermal-neutron induced fission shown in Fig.~\ref{f10} and varies between $2r_n$ and $3r_n$. A similar tendency to enhance the symmetric fission channel is observed in a highly excited system of $^{236}$U.

\subsection{Symmetric fission effect of very elongated systems}

\begin{figure}
    \centering
    \includegraphics[width=\linewidth]{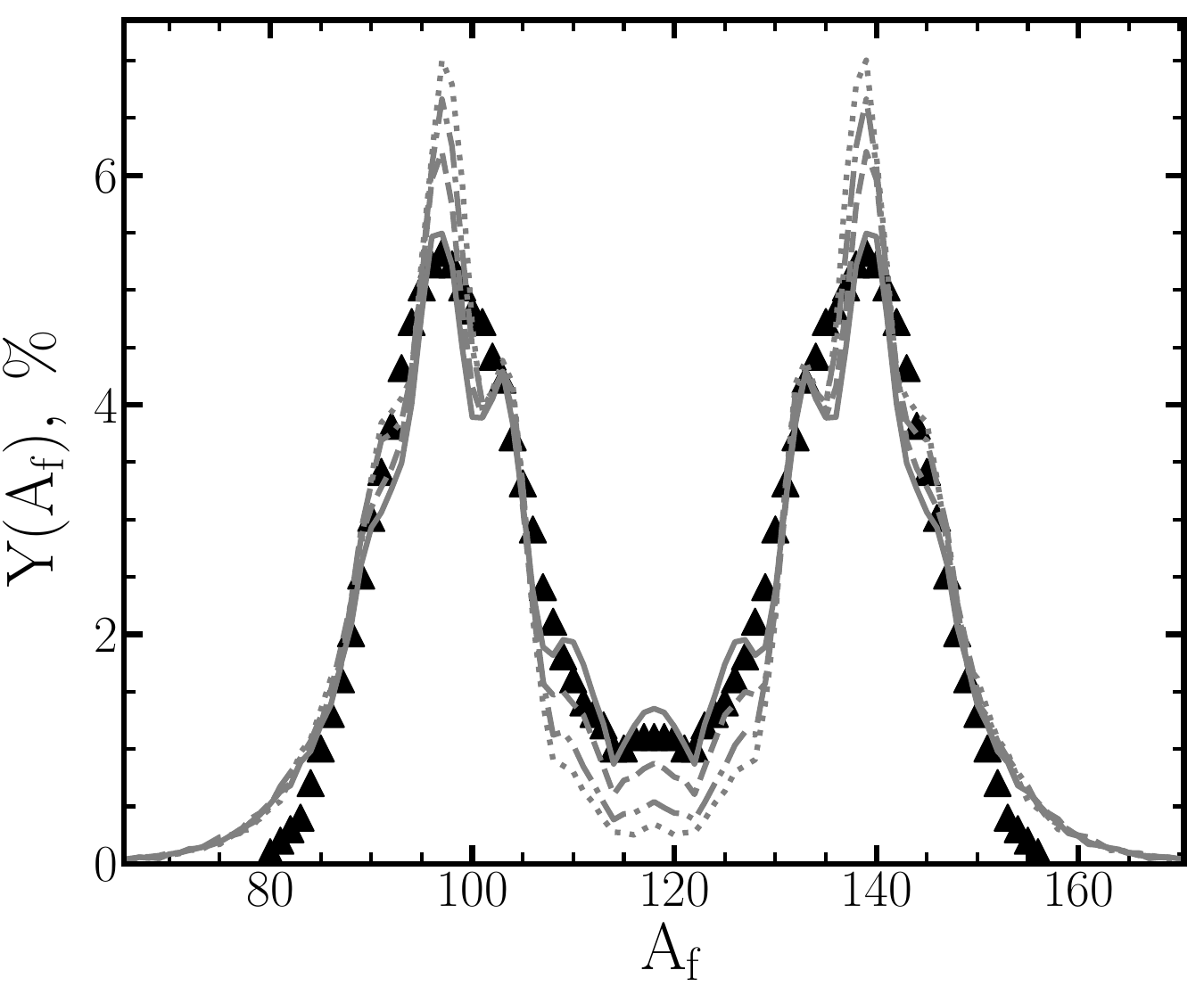}
    \vspace{-20pt}
    \caption[]{Comparison of primary FMD's for 15 MeV neutron induced fission $\rm ^{235}U$ calculated within Gaussian $P_{norm}(r_n,r_n)$ distribution of $r^{stop}_{neck}$ with the upper limit of $q^{max}_2$ is 2.35 (dotted), 2.5 (dot-dashed), 2.7 (dashed) and 2.9 (solid).}\label{f13}
\end{figure}

Now, let us return to the influence of the upper-limit value of elongation, $q^{max}_2$, on the resulting FMD's - a problem already introduced in subsection (\ref{fiss_traj}). In Fig.~\ref{f13}, the change of the FMD for medium excited fissioning $\rm ^{235}U$ system as a function of $q^{max}_2$ is presented. Recall only that if $q^{max}_2$ is continuously prolonged beyond the safe limit of $2.35$, some growing numerical uncertainties in determining the PES and the transport coefficients may appear. Therefore, the temporal evolution of this nucleus is performed until this limit is achieved. We realize that we could obtain distributions that better fit the experimental data by ignoring the fact that such inaccuracies exist. In Ref.~\cite{liu2021}, the $q^{max}_2$ value was chosen to be equal to $2.9$, which combined with $r^{stop}_{neck} = r_n$ allowed the authors to almost perfectly describe the mass and total kinetic energy (TKE) distributions for the thermal-neutron induced with an energy of 14.8 MeV fission of $\rm ^{235}U$ and $\rm ^{239}Pu$.
\begin{figure}
    \centering
    \includegraphics[width=.9\linewidth]{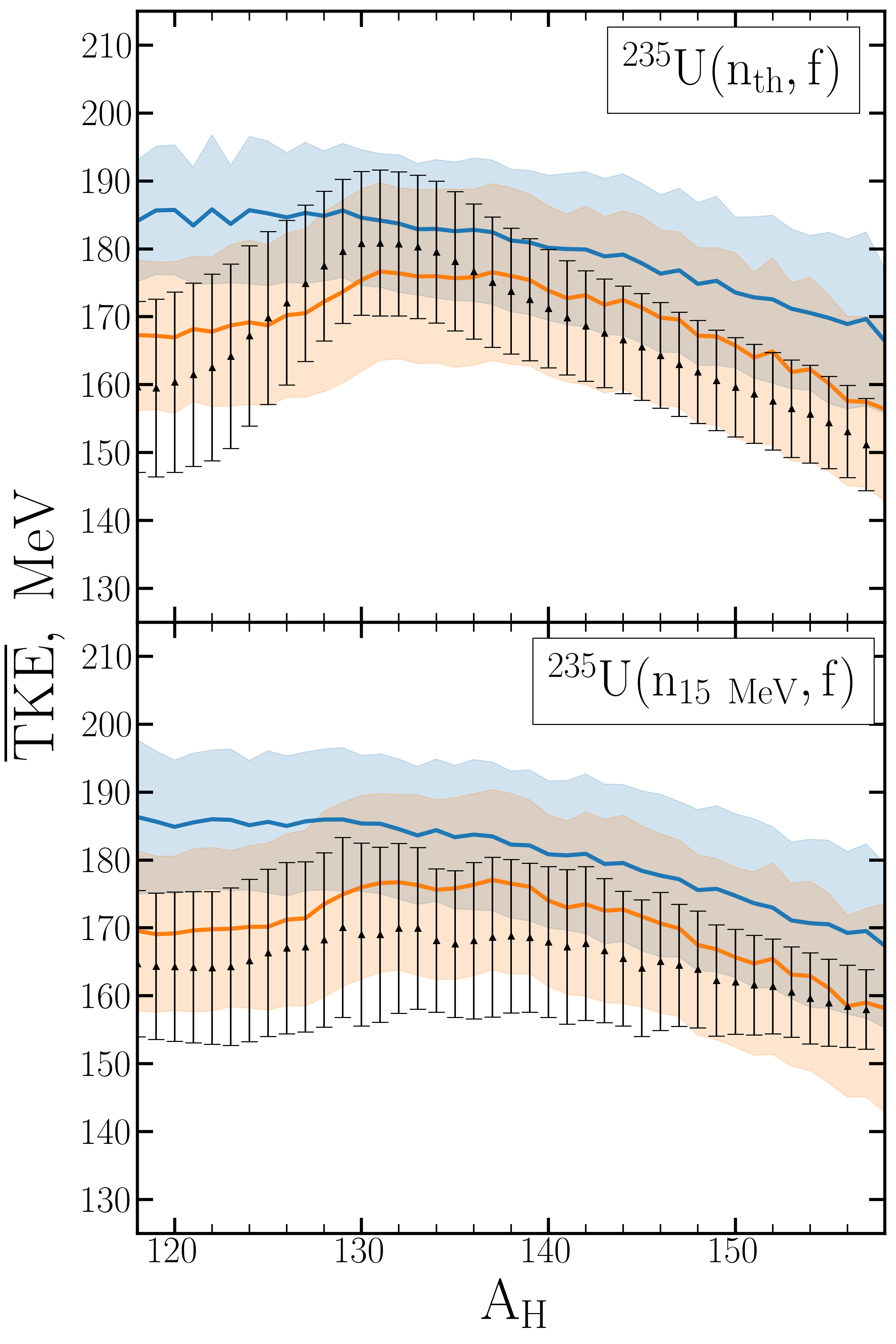}
    \vspace{-10pt}
    \caption[]{Comparison of average total kinetic energy (TKE) distributions of primary fission fragments for 15 MeV neutron induced fission of $\rm ^{235}U$ calculated within Gaussian distribution $P_{norm}(r_n,r_n)$ of $r^{stop}_{neck}$ with the upper limit $q^{max}_2$ equal to 2.35 (blue) and 2.9 (orange).}\label{f14}
\end{figure}
Figure~(\ref{f13}) proves that the distribution obtained with $q^{max}_2 = 2.8 - 2.9$ is very close to its empirical counterpart. We observe that its symmetric part grows gradually with increasing $q^{max}_2$ reaching the experimentally measured value for $q^{max}_2 = 2.9$. In addition, the height of the asymmetric peak is then perfectly reproduced. Unfortunately, the yields for extremely asymmetric mass divisions are slightly overestimated in our approach.

Using the point-charge Coulomb interaction $E_{Coul}=e^2 \frac{Z_H\,Z_L}{R_{12}}$ ($H$, $L$ denote respectively heavy and light fission fragment) supplemented by the kinetic energy of the relative motion of both fragments $\frac{1}{2} \sum_{ij} \left [\mathcal{M}^{-1} \right ]_{ij} p_i p_j$ we evaluate the average kinetic energy distributions $\bar{\mathrm TKE}$ as a function of the mass of heavy fragment, $A_H$, and present them in Fig.~\ref{f14} for $\rm ^{235}U$ nucleus. The limit value of $q^{max}_2$ fixed respectively at $2.35$ (blue curve) and $2.9$ (orange curve) are distinguished for both reactions. At the higher value of the $q^{max}_2$ limit, especially for reactions with $14.8$ MeV neutrons, the resulting distributions are noticeably closer to the experimental curves and fit well the error-bar areas.

Studying the results of Fig.~\ref{f13}, we notice that by systematic prolonging of the $q^{max}_2$ up to $2.9$, we obtain a more substantial population of the highly elongated near-symmetric yields that contribute on average to the reduction of the TKE's, particularly in part corresponding to
symmetric fragmentation.

Summarizing the above-presented test results obtained for studied uranium isotopes, we may conclude that the selection of the starting point and trajectory-termination conditions on the neck width are essential to reasonably reproduce the empirical FMD and TKE distributions, especially at higher excitation energies. The other types of constraints only allow the elimination of trajectories that are not physical, thus reducing the time to generate a necessary number of statistical samples. 
Moreover, a nontrivial relation between the final conditions and the excitation energy has been observed, the form of which has yet to be mathematically formulated. 

\section{Results and discussion}

\begin{figure}
    \centering
    \includegraphics[width=\linewidth]{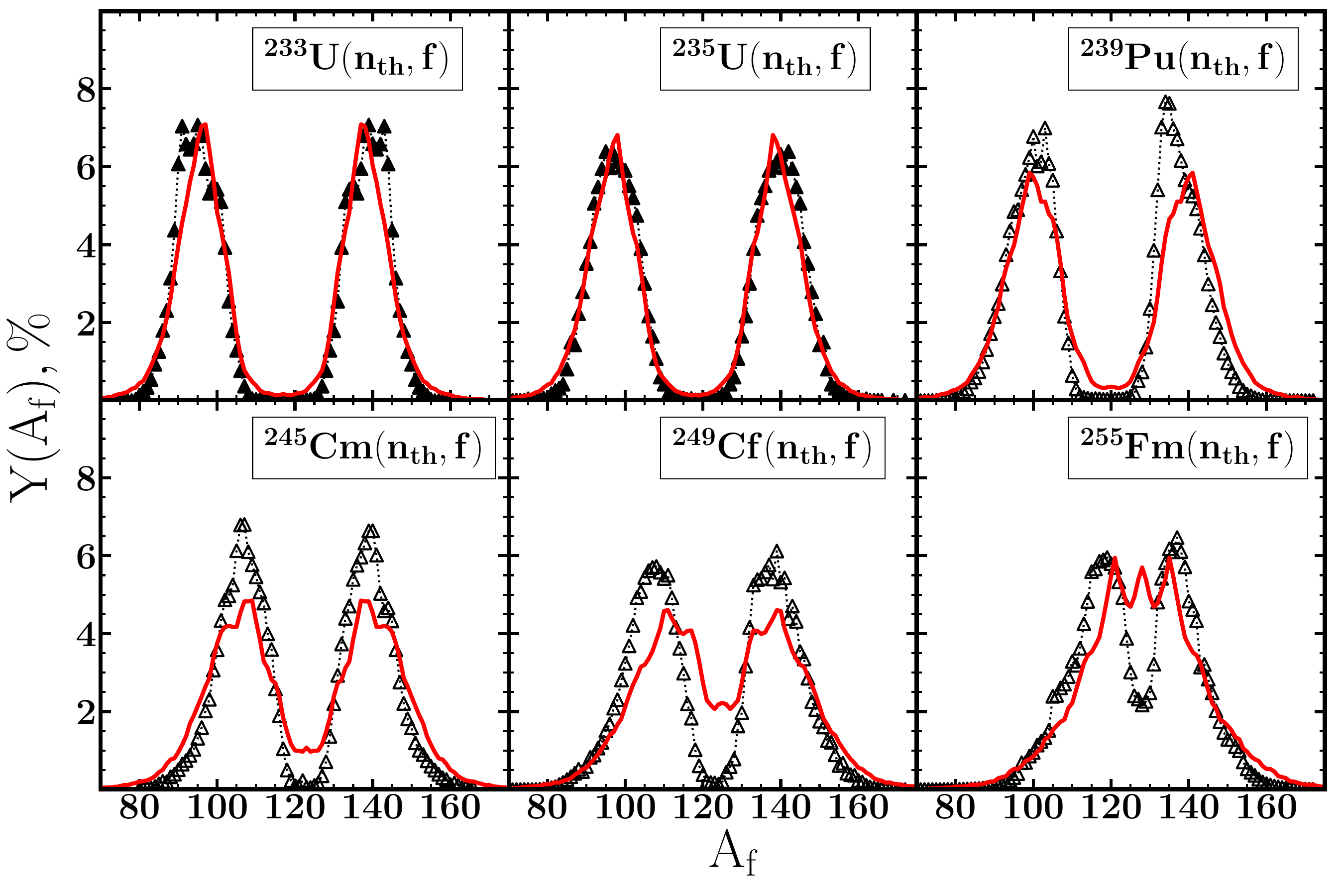}
    \vspace{-24pt}
    \caption[]{Comparison of primary FMDs calculated in our Langevin approach within 
    $P_{norm}(r_n,r_n)$ (solid red line) with primary (solid triangles) and secondary (hollow triangles) FMD's \cite{simon1990, hulet1986, schmidt_general_2016} for thermal neutron induced fission of $\rm ^{233}U$, $\rm ^{235}U$, $\rm ^{239}Pu$, $\rm ^{245}Cm$, $\rm ^{249}Cf$ and $\rm ^{255}Fm$ nuclei.}\label{f15} 
\end{figure}

After establishing the initial and termination (final) criteria for Langevin trajectories by studying the fission of $\rm ^{235}U$ isotope as the benchmark case, let us broaden the applicability of the model in question to the other actinide elements. In particular, we will focus on the isotopes of $\rm ^{233}U$, $\rm ^{239}Pu$, $\rm ^{245}Cm$, $\rm ^{249}Cf$ and $\rm ^{255}Fm$, for which experimental data on FMD are available. Obtained distributions for chosen nuclei from among these isotopic chains are illustrated in Figs~\ref{f15}. The evaluated results agree with experimental data in medium-heavy actinides, such as uranium, plutonium, and curium. Nevertheless, some larger discrepancies between estimated and empirical distributions are present in selected heavy actinides of californium and fermium isotopes. Although the available experimental data refer to the distributions of secondary fragments (after the emission of light particles from compound nucleus as well as from fission fragments), the mutual shift of both those distributions by a couple of mass units for $\rm ^{250}Cf$ or appearing of the symmetric-fission peak in our FMD's of $\rm ^{255}Fm$ cannot be fully explained by the effects of light particle evaporation alone.
Also, as commonly known, in Cf and Fm nuclei, a rapid transition from the dominant asymmetric to symmetric fission mode, caused by adding two neutrons, is noticed. The reproduction of that is a particular challenge for our model. 
Recall that spontaneous or induced fission processes are probabilistic phenomena associated with overcoming the fission barrier between the ground state or some excited state and an exit point of the same energy, by definition, located outside the barrier. In a quantum approach, the probability of passing the barrier is crudely dependent on the barrier shape and the number of hits on the barrier per time unit. In contrast, the barrier is not "tunneled" in our Langevin-like semi-classical approach. However, it must be over-jumped by a system with kinetic energy greater than the barrier height in the initial evolution stage.
 \begin{figure*}
    \centering
    \includegraphics[width=\linewidth]{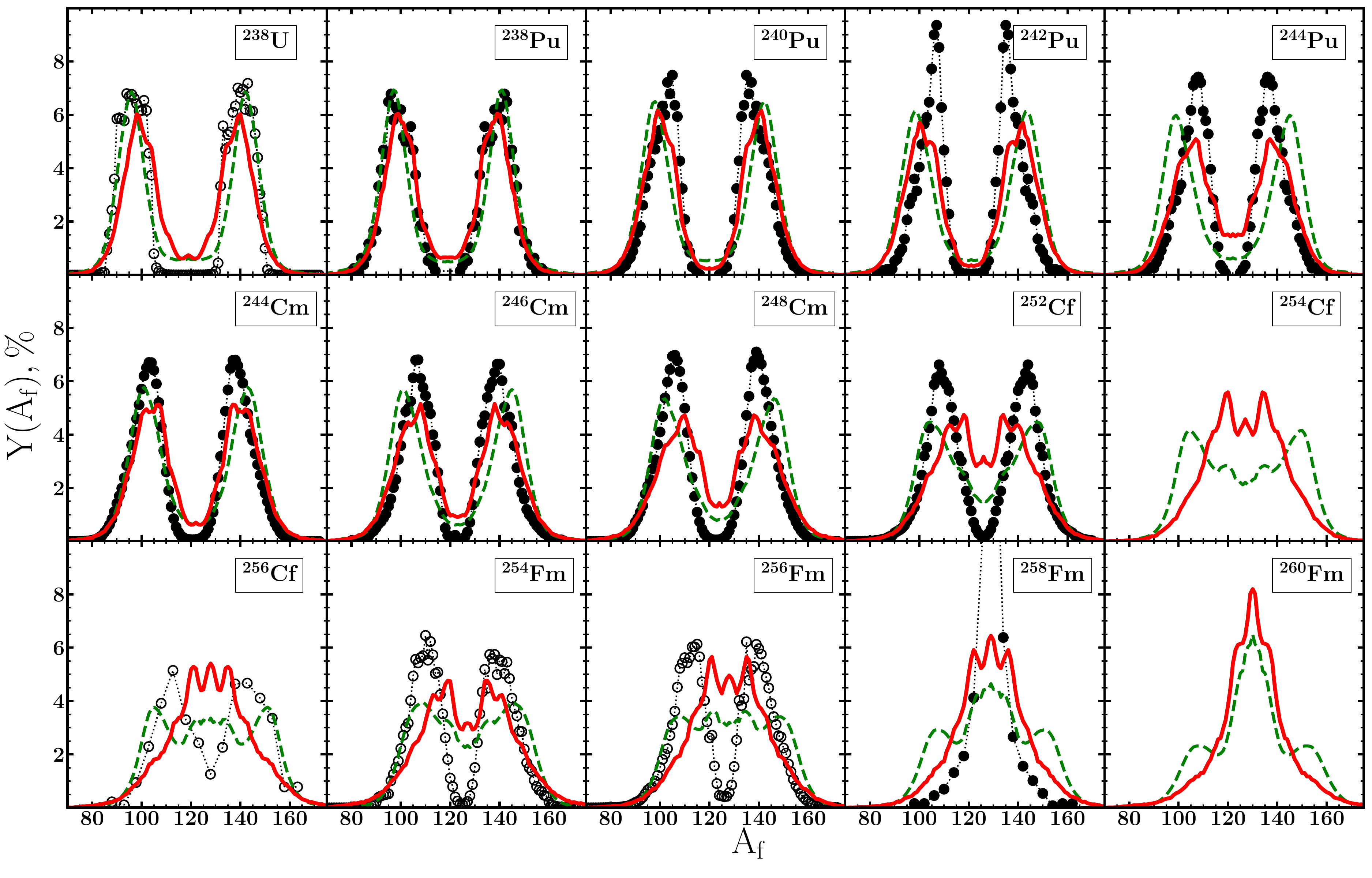}
    \vspace{-12pt}
    \caption[]{Comparison of primary FMD's (solid red line) calculated in Langevin approach within $P_{norm}(r_n,r_n)$ with experimental \cite{dematte1997, pleasonton1973, hoffman_1980, hulet1986, schmidt_general_2016} primary (solid circles) and secondary (hollow circles) FMDs  for spontaneous fission of $\rm Pu$, $\rm Cm$, $\rm Cf$ and $\rm Fm$ nuclei also calculated with FMDs calculated within BOA method~\cite{pomorski2021}(green dashed)} \label{f16}
\end{figure*}

Using the method of determining the starting points, given by formula~\eqref{eq:orig_cond}, which are located slightly beyond the outer saddle point, we perform the calculations of Langevin trajectories corresponding to the spontaneous fission for the following even-even nuclei series: $\rm ^{238}U$, $\rm ^{238-244}Pu$, $\rm ^{244-248}Cm$, $\rm ^{252-256}Cf$ and $\rm ^{254-260}Fm$. The trajectory evolution is initiated using similar rules as in the case of induced fission, described in the previous sections. A particular value of $r^{stop}_{neck}$ for which a given trajectory is terminated at the pre-scission point is drawn at the beginning of each trajectory with the Gaussian probability distribution, $P_{norm}(r_n,r_n)$, as already used in the study of $\rm ^{235}U$ isotope. Figure~\ref{f16} illustrates the final FMDs for the spontaneous fission of mentioned nuclei. At the background of generally satisfactory agreement between theoretical and experimental, mainly primary distributions of mass fragments in these nuclei, we note in $\rm ^{252}Cf$ that although the evaluated distributions reproduce the dominance of asymmetric yields, it considerably overestimate the number of symmetrical fragments and, in addition, are much too narrow. Notice that in $\rm ^{254,256}Fm$ $\rm ^{256}Cf$, the measured distributions include the effect of light-particle emission. Both the compared in this figure distributions, however, differ radically. 

Searching the discrete grids of the potential energy of californium and fermium nuclei with neutron numbers corresponding to the transition area from asymmetric to symmetric FMD, we can find more than one point describing possible configurations of the exit from under the barrier. We, therefore, postulate that each such state should be treated as the starting point to perform the Langevin fission simulation. The final FMD thus obtained, say "partial fragment mass distributions," should be superimposed with appropriate weights to obtain the final FMD. A classical measure of these weights may be the values of the action integrals evaluated between a given starting and exit point. The latter may be found either in the symmetric or asymmetric fission valley. This approach can be used mainly for Cf and Fm nuclei, in which the system decides where to go within a small bifurcation area after crossing the barrier. Nevertheless, this issue is beyond the scope of this work and will be addressed in future investigations.

For a complete comparison, we also include the experimental distributions of primary and secondary fission fragments, as well as the evaluated distributions obtained within the $P_{norm}(r_n,r_n)$ neck-radius normal distribution, and those calculated within the framework of the static Born-Oppenheimer (BOA) model. This latter approach is based on an approximate solution of the eigenvalue problem of the three-dimensional collective Hamiltonian. A more detailed description of that and the corresponding results for a wide range of even-even actinides can be found in~\cite{pomorski2017,pomorski2021}. Despite the different theoretical underpinnings of these two models, they both exploit identical PES's and inertia parameters associated with our three-dimensional Fourier deformation space.

\section{Summary}

This work presents a quasi-classical dynamical approach to simulate the stochastic nature of the fission of a compound nucleus using a system of Langevin equations which needs as the input the
free (Helmholtz) energy based on the macroscopic-microscopic PES. The calculations are done in the space of three relevant for the fission process Fourier surface deformations to describe the nucleus elongation, mass asymmetry, and neck thickness. As widely known, the impact of the non-axiality degree of freedom on the PES is irrelevant beyond the outer saddle and in the neighborhood of the scission configurations, thus, is neglected in this study. Such a simplification allows generating of hundreds of thousands of Langevin trajectories for a single nucleus within a reasonable time of tens of minutes.

We emphasize the importance of initial and trajectory-termination conditions, which are independent of the particular realization of the Langevin framework. This last condition appears particularly important as it defines the "critical" width of the neck that allows the composite nucleus to be divided into fragments.
Since our calculations are carried out on a finite PES grid, we pay particular attention to the exact setting out of the grid boundaries to capture all essential fission modes, predominantly strongly asymmetric, and additionally not to consider non-physical energy configurations showing up for considerably large elongations.

After analyzing the trajectory termination condition, we concluded that noticeably better results could be obtained if a normal probability distribution $P_{norm}(r_n,r_n)$ of the size of the neck radius instead of its fixed value is used. The maximum of $P_{norm}$ is located at the mean value of the nucleon radius and the exact value of $r_n$ as its standard deviation. 
With the increasing temperature of the system, the value of $r^{stop}_{neck}$ is shifted towards larger values.

The results on the induced and spontaneous fission of various actinide nuclei, such as $\rm U$, $\rm Pu$, $\rm Cm$, $\rm Cf$, and $\rm Fm$ show generally good agreement with experimental data for medium and some selected heavy actinides. However, there is a discrepancy in the overall behavior of FMD in the $\rm Cf$ series. We strongly hope that applying the concept mentioned above of the superposition of various partial FMD's initiated from a different available exit from the barrier points and using modified Fourier-over-spheroid shape parametrization~\cite{pomorski2023} would allow us to better reproduce the mentioned effect of the abrupt transition between asymmetric to symmetric fragmentation.
Further work is needed to improve the model's ability to describe other fission characteristics, such as secondary FMD's and TKE's corrected by the effect of light particle evaporation.

\vspace{-6pt}
\begin{acknowledgments}
\vspace{-6pt}

The authors are grateful to K. Pomorski and C. Schmitt for their support and fruitful discussions. This work is also partly supported by the Polish National Science Center by SHENG-1 project (Grant No 2018/30/Q/ST2/00185) and NAWA STER project "UMCS Doctoral Schools - Your Success in Globalized World of Science" (Grant No BPI/STE/2021/1/00006/U/00001)
\end{acknowledgments}

\bibliography{references}

\end{document}